\documentclass[10pt,a4paper,aps,prd,twocolumn,showpacs,superscriptaddress]{revtex4-1} 
\usepackage[english]{babel}

\usepackage[utf8]{inputenc}
\usepackage[T1]{fontenc}
\usepackage{url} 
\usepackage{hyperref}
\usepackage{amsmath}
\usepackage{slashed} 
\usepackage{graphicx}

\usepackage{bm} 
\usepackage{mathtools,mathrsfs}
\usepackage{amsmath,amssymb}
\usepackage{bbold}
\usepackage{bm,bbm}
\usepackage{siunitx}
\usepackage[capitalise]{cleveref}
\usepackage[normalem]{ulem}

\usepackage{physics}
\usepackage{xcolor}

\newcommand{\be}{\begin{equation}}
\newcommand{\ee}{\end{equation}}
\newcommand{\bi}{\begin{itemize}}
\newcommand{\ei}{\end{itemize}}
\newcommand{\vphi}{\varphi}
\newcommand{\trm}[1]{\textrm{#1}}
\newcommand{\mbf}[1]{\mathbf{#1}}
\definecolor{bkcol}{rgb}{0.0, 0.6, 0.9}

\begin{document}
\allowdisplaybreaks

\title{High-resolution modeling of nonlinear Compton scattering in focused laser pulses}
\author{C. F. Nielsen}
\affiliation{Department of Physics and Astronomy, Aarhus University, Denmark}
\author{R. Holtzapple}
\affiliation{Department of Physics, California Polytechnic State University, San Luis Obispo, California 93407, USA}
\author{B. King}
\affiliation{Deutsches Elektronen-Synchrotron DESY, Notkestr. 85, 22607 Hamburg, Germany}
\affiliation{Centre for Mathematical Sciences, University of Plymouth, PL4 8AA, UK}
\date{\today}
\begin{abstract}
A semi-classical approach is used to calculate radiation emission in the collision of an electron with an intense focused laser pulse. The results are compared to predictions from the locally constant field and locally monochromatic approximations. It is found that simulations employing the semi-classical approach capture features in the energy spectra, such as subharmonics and bandwidth structure, which are beyond local approaches. The formation length is introduced as a diagnostic to select between approaches as the electron is propagated through the pulse.
\end{abstract}
\maketitle

\section{Introduction}
The radiation produced in the nonlinear Compton scattering of high energy electrons and positrons moving in strong electromagnetic fields has been \cite{burke_positron_1997,bula_observation_1996} and is being intensively studied \cite{poder_experimental_2018,cole_experimental_2018,gonsalves_pw_8gev_2019,leemans_multi-gev_2014,wang2013quasi,kim2013enhancement,tanaka_current_2020,turcu_strong_2015,maksimchuk2020zeus,cartlidge_light_2018,weber_p3_2017,leemans2010berkeley}. The prevalent theoretical approach is the background field method or `Furry' picture \cite{Furry:1951zz}: the interaction between charges and the background is determined by solving equations of motion in that background exactly, whereas radiated fields are included perturbatively (some reviews of strong-field QED can be found in \cite{ritus_1985,2009RPPh...72d6401E,AntoninoReview,Narozhny:2015vsb,Fedotov:2022ely}). The resulting scattering matrix elements depend in a highly nonlinear way on outgoing momenta, which means that calculation of inclusive probabilities, even for lowest-order tree-level processes, is non-trivial. Therefore, approximation frameworks have been developed, which implement QED Monte Carlo generators in numerical simulation of intense laser-matter interaction (some reviews can be found in \cite{Elkina:2010up,Gonoskov:2014mda,RIDGERS2014273,Lobet2016,Gonoskov:2021hwf}). A central approximation for including QED effects in these simulations, is the locally constant field approximation (LCFA), which has been benchmarked against numerical evaluation of analytical results in numerous studies \cite{BenKing2015,DiPiazza2018,Aleksandrov:2018zso,DiPiazza2019,BenKing2019,Raicher:2020nkq,Torgrimsson:2020gws}. However, the experiments E320 \cite{E320Talk} at FACET-II and LUXE \cite{Abramowicz:2021zja} at DESY, will measure strong-field QED effects in the intermediate intensity regime, outside the region of applicability of the LCFA. Furthermore, these experiments will employ a conventionally-accelerated electron beam and therefore can measure strong-field QED effects at a higher accuracy than has so far been possible using laser-wakefield accelerated beams \cite{cole_experimental_2018,poder_experimental_2018}. In response to these developments, the locally monochromatic approximation (LMA), which was originally employed in the E144 experiment in the mid-90s \cite{bula_observation_1996,burke_positron_1997,bula97,bamber.prd.1999} has recently been more intensively studied. The LMA has been derived from QED and formalized \cite{Heinzl:2020ynb}, investigated analytically \cite{King:2020hsk,Tang:2021qht}, included in simulation frameworks \cite{cain,hartin.ijmpa.2018,ptarmigan22} and benchmarked with numerical evaluation of exact formulas \cite{Torgrimsson:2020gws,Blackburn:2021cuq,Blackburn:2021rqm,Tang:2022tmn}. (The open-source code Ptarmigan \cite{ptarmigan22} employs the LMA and is being used to simulate the interaction point of the LUXE experiment.) The LMA trades the versatility of the LCFA with accuracy in reproducing harmonic structure, by assuming a background with two well-defined timescales (e.g. pulse envelope and carrier frequency). The given QED process is calculated exactly over the short timescale, whereas changes in the slow timescale are included adiabtically \cite{Narozhnyi:1996qf,McDonald:1997,seipt.pra.2011,seipt.jpp.2016}. These local approaches take into account interference on different length scales: the LCFA on the sub-wavelength scale and the LMA on the wavelength scale (the `local' expansion is performed in the pulse envelope). In contrast to these local approaches is an alternative  based on the fact that in a plane-wave background, the Dirac equation is semi-classical exact: electrons and positrons can be modeled as point-like. In the semi-classical approach, one evaluates the spectrum numerically as an integral over the particle trajectory through an \emph{arbitrary} electromagnetic field \cite{baier1968quasiclassical,Matveev1957,kimball_1986}. Depending on how much of the trajectory is integrated over, this method can take into account interference on the length scale of the entire pulse. (As indicated above, in a plane wave background, this approach is even exact.) The semi-classical approach has been extended analytically and applied to focused laser backgrounds \cite{DiPiazza:2016maj,DiPiazza:2016tdf,DiPiazza:2020wxp} as the leading order WKB result, which assumes the hierarchy \cite{DiPiazza:2013vra}: $m\ll ma_{0} \ll \epsilon$, where $m$ is the electron mass, $\epsilon$ the particle energy and $a_0$ the intensity parameter, to be introduced below. More recently it has also been applied to standing-wave backgrounds \cite{Lv:2021ayt}. The numerical method has been applied to the case of high energy electrons and positrons propagating through the non-plane-wave fields of oriented single crystals, and shown to give reliable results that agree well with experiment \cite{Nielsen2020LL,NielsenGPU2020}. We call the semi-classical formulation in terms of radiation integrals that we use in this paper, the Belkacem, Cue, Kimbal (BCK) model. Although analytical expressions for the BCK model have been known for decades, the difficulty of evaluating the radiation spectra has prevented a detailed application to electron-laser collisions.

In the following, we compare what the LCFA, LMA and BCK models predict for the radiation spectrum generated by the collision of electrons with an intense focused laser pulse.  We assume the laser pulse is a \emph{thin target} i.e. that electrons only undergo \emph{single} nonlinear Compton scattering. With several upcoming experiments aiming to use these models for theoretical predictions, a detailed investigation of the different radiation models is useful in assessing their accuracy. In Sec. II, we highlight some relevant points in deriving the theoretical models and in Sec. III we discuss the concept of `formation length' which is useful in determining when a model will be sufficiently accurate or not. In Sec. IV we present results of benchmarking the models: first, in a monochromatic background; then in a finite plane wave pulse, which is compared against direct numerical evaluation of the QED expression; then, in a focused laser background. In Sec. V, a comparison of the computation time for each model is made and in Sec. VI the results are summarized and the paper concluded.

\section{Theoretical formalism}
The common starting point for all three theoretical models used for radiation emission investigated in this paper is the transition matrix element $M$, for single nonlinear Compton scattering
\cite{ritus_1985,BERESTETSKII1982,baier1968quasiclassical} that contains the transition matrix element $M$.  The transition matrix element is given by 
\begin{equation}\label{MatrixElement}
    M = -ie\int \bar{\Psi}_{p'}(x)\hat{e}'^{\ast}\Psi_p(x)\frac{\text{e}^{ik'\cdot x}}{\sqrt{2k_0'}}\text{d}^4x ,
    \end{equation}
for an electron that transitions from an initial state $\Psi_{p}(x)$ to a final state $\bar{\Psi}_{p'}(x)$ (where $\bar{\Psi}_{p'} = \Psi^{\dagger}_{p'}\gamma_{0}$) and radiates a single photon (volume factors have been set equal to unity). (We use the shorthand $k'\cdot x =k'_{\mu}x^{\mu}$, use $\hat{\phantom{}}$ to denote an operator and set the speed of light in vacuum $c=1$, but keep factors of the reduced Planck constant, $\hbar$).
In \cref{MatrixElement} $e<0$ is the electron charge, $\hat{e}'$ is the photon polarization vector, $p$ and $p'$ is the electron four-momentum in its initial and final state respectively and $k'$ is the photon four-momentum in the final state. 

The radiation models investigated in this work are differentiated by how the squared transition matrix element is approximated. Derivation of the local approximations is based on the exact solution to the Dirac equation in a plane wave background (so-called Volkov wavefunction \cite{volkov1935class,ritus_1985,BERESTETSKII1982}), and the generalization of local approximations in simulations to non-plane-wave backgrounds follows by calculating the relevant local momentum in a non-plane-wave context. The vector potential of the background, $A^\mu(\phi)$, is assumed only to depend on the invariant variable $\phi = k \cdot x$. Here $k_\mu$ is the zero-length wavevector of the background field with $k_\mu = (|\bold{k}|,\bold{k})_{\mu}$ where $\bold{k}$ determines the propagation direction of the external field. The Volkov wavefunction can be written as:
\begin{equation}
    \Psi_p(x) = \left[1+\frac{e\slashed{k}\slashed{A}(\phi)}{2k\cdot p}\right]\frac{u(p)}{\sqrt{2p_0}}\text{e}^{iS/\hbar},
\end{equation}
where $u(p)$ is the  free particle plane wave bispinor. The parameter $S$ is the classical action of an electron moving in an electromagnetic field with vector potential $A^\mu(\phi)$ given by
\begin{equation}
    S = -p\cdot x - e\int^{\phi}_0\left[\frac{p\cdot A(\phi')}{k\cdot p}-\frac{eA(\phi')^2}{2k\cdot p}\right]\text{d}\phi'.
\end{equation}
The electron wave function has a classical action, which is a reflection that the Volkov wave function is semi-classical \cite{ritus_1985}. This allows the motion of the electron in the external electromagnetic field to be treated classically. In contrast, the interaction between the electron and the radiated photon is treated quantum mechanically. This is true for all radiation models investigated in this paper. The Volkov wave functions are solutions to the Dirac equation, consequently they are completely quantum mechanical, therefore no semi-classical characteristics are inferred. The semi-classical nature of the wave function could have been predicted by looking at the quantum effects in the photon emission process and in the trajectory of the electron moving in an electromagnetic field. In non-plane-wave fields, quantum effects in the \emph{trajectory} of the particle become important when the energy levels in the electromagnetic field become comparable to the energy of the particle. One example where this can be clearly seen is in the propagation of a particle perpendicular to a constant magnetic field $H$, which is considered in, e.g. refs. \cite{BERESTETSKII1982, baier1968quasiclassical}. In this background, the energy levels of the electron are quantized with an energy spacing of
\begin{equation} \label{}
    \hbar\omega_0 = \hbar \frac{v|e|H}{|\bold{p}|}  \approx \hbar\frac{|e|H}{\epsilon},
\end{equation} 
where $\omega_0$ is the classical frequency of revolution of a particle with energy $\epsilon$, momentum $\bold{p}$, and speed $v$. The approximation in Eq. (4) holds for $\gamma \gg 1$ with $\gamma$ being the relativistic Lorentz factor. The motion is classical when $\hbar\omega_0/\epsilon \ll 1$.
It is convenient to introduce the quantum nonlinearity parameter $\chi$ and the critical Schwinger field $H_0$ \cite{BERESTETSKII1982}
\begin{equation}
    \chi=\sqrt{-(F_{\mu\nu}u^\nu)^2}/H_0, \;\;\; H_0=m^2/|e|\hbar \approx 4.4 \cdot 10^{13}\;\text{G} \; ,
\end{equation}
where $F^{\mu\nu}$ is the external electromagnetic field tensor, $u^{\mu}$ is the four-velocity of the electron. The parameter $\chi$ can be expressed as the ratio between the Lorentz boosted electromagnetic field and the critical field in the rest frame of the particle. The trajectory is considered to be classical when
\begin{equation}
    1 \gg \frac{\hbar\omega_0}{\epsilon} = \frac{vmH}{\gamma H_0}  \approx\frac{H}{H_0}\frac{1}{\gamma^2} \approx \frac{\chi}{\gamma^3}. \label{eqn:ClassTrajWhen1}
\end{equation} 
Quantum effects in the \emph{emission process} (i.e. electron recoil) become important when the energy of a single emitted photon is comparable to the energy of the emitting particle. Classically, a charged relativistic particle in an electromagnetic background can emit photons via synchrotron emission with frequencies up to the critical frequency $\omega_c\approx \omega_0\gamma^3$ \cite{jackson_classical_1998}. For a relativistic particle moving perpendicular to a magnetic field, the condition for the emission process to be classical is
\begin{equation}
   1 \gg \frac{\hbar \omega_c}{\epsilon} \approx \gamma\frac{H}{H_0} \approx \chi.
\end{equation}
From the above discussion, it is evident that the parameter $\chi$ is important in determining if the emission process or the trajectory can be treated classically or has to be treated quantum mechanically. 
(Presently, no experiment has been close to producing fields in the lab frame comparable to the Schwinger critical field, otherwise these arguments would also have to be generalized to also consider the magnitude of the electromagnetic invariants.)
Due to the extra $\gamma^3$ factor on the radiation condition, quantum effects in the radiation process quickly start becoming important as the energy increases (these have been noted in plane-wave backgrounds in the phenomena of e.g. straggling \cite{Blackburn:2014cig} and quenching \cite{Harvey:2016uiy}). We see that \cref{eqn:ClassTrajWhen1} provides justification for treating the trajectory of the electrons classically, even though a quantum theory is being used.

In addition to the question of `classical' vs. `quantum', there is also the question of \emph{coherence}. Given a fixed point on the trajectory, over what length does the amplitude for emission add coherently with neighboring points on the trajectory? This is sometimes referred to as the `interference' or `coherence' length. This can be understood by considering the probability for emission, which is proportional to the mod-square of $M$ (see \cref{MatrixElement}). The only non-trivial space-time integrals are over variables $t$ and $t'$. It is useful to define:
\be
\tau = \frac{t+t'}{2};\quad  \Delta t = t-t',
\ee
where $\tau$ is the average time co-ordinate, and $\Delta t$ is the `interference' time co-ordinate. Then the probability can be cast as an integral over the average position of the electron, quantified by $\tau$ and an interference window in $\Delta t$ around $\tau$. Depending on the particle energy and field strength, the integral in $\Delta t$ can be approximated in different ways.

In the following sections we outline the derivations of three models to evaluate the radiation spectrum from nonlinear Compton scattering of an electron moving in an electromagnetic background, with the goal of highlighting the major differences between the methods. This will provide an understanding of the advantages and limitations of each model, which will be investigated later.

\subsection{Locally Monochromatic Approximation}
This section outlines some of the key points in deriving the LMA in a linearly polarized potential. It differs from \cite{Heinzl:2020ynb} in that only the diagonal terms in the double harmonic sum are kept (the off-diagonal terms include an integral over an oscillating phase and have so far not been reported as having any significant impact on energy spectra). Consider an electron moving in a linearly polarized plane wave field with a vector potential of the form $A^\mu(\phi) = a^\mu \text{cos}(\phi)$, where $a^\mu$ is a constant four-potential amplitude.  Evaluating the classical action integral analytically results in the following wave function \cite{Nikishov1964,Brown:1964zzb,ritus_1985}
\begin{equation}
\begin{aligned}\label{WaveNLC}
    \Psi_p(x) = \left[1+\frac{e\slashed{k}\slashed{a}\text{cos}(\phi)}{2k\cdot p}\right]\frac{u(p)}{\sqrt{2q_0}}&
    \\\text{exp}\left(-i\frac{ea \cdot p}{k\cdot p}\text{sin}(\phi)+\right.i\frac{e^2a^2}{8k\cdot p}&\text{sin}(2\phi)-iq\cdot x \!\!\!\left.\phantom{\frac{}{}}\right),
\end{aligned}
\end{equation}
where $q^\mu = p^\mu-k^\mu e^2a^2/4k\cdot p$, is the quasimomentum of the electron in the plane wave (the normalization in terms of $q^{0}$ follows the arguments in \cite{ritus_1985}). It is evident that the transition matrix element in \cref{MatrixElement} will depend on functions of the type
\begin{equation}
    \text{cos}^n(\phi)\text{e}^{i(\alpha\text{sin}(\phi)-\beta\text{sin}(2\phi))}, \;\;\; n = 0,1,2,
    \label{cosinefuncs} 
\end{equation}
with the parameters 
\begin{equation}
    \alpha = e\left(\frac{a\cdot p'}{k\cdot p'}-\frac{a\cdot p}{k\cdot p}\right), \;\; \beta = \frac{e^2a^2}{8}\left(\frac{1}{k\cdot p'}-\frac{1}{k\cdot p}\right).
    \label{alphabeta}
\end{equation}
The functions in \cref{cosinefuncs} can be written in terms of a discrete Fourier series
\begin{equation}
        \text{cos}^n(\phi)\text{e}^{i(\alpha\text{sin}(\phi)-\beta\text{sin}(2\phi))} = \sum_{s = -\infty}^\infty A_n(s,\alpha,\beta)\text{e}^{is\phi}, \label{eqn:Four1}
\end{equation}
where $s$ is an integer and 
\begin{equation}
    A_n(s,\alpha,\beta) = \frac{1}{2\pi}\int_{-\pi}^\pi \cos^n(\phi)\text{e}^{i(\alpha\text{sin}(\phi)-\beta\text{sin}(2\phi)-s\phi)}\text{d}\phi.
\end{equation}
The function $A_{0}(s,\alpha,\beta)$ is related to the `generalized Bessel function', $J_{s}(\alpha,\beta)$ where \cite{PhysRevE.79.026707}:
\be
A_{0}(s,\alpha,\beta) = \sum_{m=-\infty}^{\infty} J_{2m+s}(\alpha)J_{m}(\beta) :=J_{s}(\alpha,\beta),
\ee
and $A_{1}$ and $A_{2}$ can be related to the generalized Bessel function by expanding out the cosine in exponentials.

The parameters in the above expression are given by
\begin{align}
    \alpha &= z \cos(\theta),\\
    z &= \frac{a_0^2\sqrt{1+a_0^2/2}}{\chi}\sqrt{u(u_s-u)}, \label{eqn:z1}\\
    u_s &= \frac{2 s \chi}{a_0(1+a_0^2/2)},\\
    \beta &= \frac{a_0^3 u}{8\chi},
\end{align}
where $u = k\cdot k'/(k\cdot q') \approx \hbar\omega /(\epsilon-\hbar\omega) $ is the ratio of photon and outgoing electron lightfront momenta, related to the ratio of energies, and $\theta$, defined between $0$ and $2\pi$, is related non-trivially to the angle between the process constituents (see e.g. \cite{ritus_1985}). The parameter 
\begin{equation}
    a_0 = \frac{|e| \sqrt{-a^2}}{m},
\end{equation}
is an invariant often called the classical nonlinearity parameter. The parameter $a_{0}$ is equivalent to the work done over a Compton wavelength in units of the field's central photon energy.  (In some literature the root-mean-squared value of the field strength is used instead of $\sqrt{-a^2}$ \cite{E144}, then one must include a factor $1/\sqrt{2}$ and the resulting cross sections altered accordingly.)

Employing the wave function in \cref{WaveNLC} and the Fourier expansion in \cref{eqn:Four1} in  \cref{MatrixElement} and integrating over $\text{d}^4x$ yields a delta function  $\delta(sk_\mu+q_\mu-q_\mu'-k_\mu')$ in the matrix element, $M$, for each term in the sum over $s$. 

The resulting probability for an electron moving in a linearly polarized plane wave to emit a photon with energy ratio $x_\omega = \hbar\omega/\epsilon$ per unit time can be written 
\begin{equation}
\begin{aligned}
        \frac{\text{d}P}{\text{d}t\text{d}x_\omega} = \frac{\alpha m^2}{2\pi q_0}& \sum^\infty_{s=1}\int^{2\pi}_0\left(-A_0^2\right.+\\ &\left.a_0^2\left[1+\frac{u^2}{2(1+u)}\right](A_1^2-A_0A_2)\right)\text{d}\theta,
        \end{aligned}\label{eqn:LMA1}
\end{equation}
where the arguments of the $A_{n}$ functions have been suppressed and where the ratio $x_\omega$ relates to $u$ by ${u = x_\omega/(1-x_\omega)}$. The rate in \cref{eqn:LMA1} is in the same form as used to model the E144 experiment \cite{bamber.prd.1999} and agrees exactly with the recent derivation from QED \cite{Heinzl:2020ynb}, if the diagonal terms in the double harmonic sum are neglected and the intensity $a_{0}$ in the arguments of $A_{n}$ is localized to the envelope value of the potential. We refer to \cref{eqn:LMA1} as the LMA in this paper. To apply the LMA to simulations, all instances of the electron momentum can be replaced with local cycle-averaged quasimomentum \cite{Blackburn:2021cuq}.

\subsection{Locally Constant Field Approximation}
It was shown in \cite{Heinzl:2020ynb} that if the limit $a_{0}\to\infty$ is taken, whilst holding $\chi$ constant (e.g. by taking the limit $\omega_{l}\to 0$), the LMA tends to the LCFA just as the monochromatic result tends to the constant crossed field result in the same limit \cite{ritus_1985}. Here we recap arguments from  \cite{ritus_1985} how the rate in a constant crossed field is calculated. This can be achieved by immediately setting $A^\mu(\phi) = a^\mu \phi$ at the beginning of the calculation so the wavefunction becomes:
\begin{equation}
\begin{aligned}
    \Psi_p(x) = \left[1+\frac{e\slashed{k}\slashed{a}\phi}{2k\cdot p}\right]\frac{u(p)}{\sqrt{2p_0}}&
    \\\text{exp}\left(\!\phantom{\frac{}{}}\right.-i\frac{ea\cdot p}{2k\cdot p}&\phi^2+i\frac{e^2a^2}{6k\cdot p}\phi^3-ip\cdot x \!\!\!\left.\phantom{\frac{}{}}\right).
\end{aligned}
\end{equation}
To make progress one can, for example, Fourier-transform the nonlinear part of the exponent of the transition matrix elements, which leads to the introduction of functions $C(s,\alpha,\beta)$, defined through:
\begin{equation}
    \phi^n\exp\left[i\left(\frac{\alpha\phi^2}{2}-\frac{4\beta\phi^3}{3}\right)\right],\;\;\; n = 0,1,2,
\end{equation}
where $\alpha$ and $\beta$ are defined in \cref{alphabeta}. The functions can be written in terms of a continuous Fourier integral 
\begin{equation}
     \phi^n\text{e}^{\left[i\left(\alpha\phi^2/2-4\beta\phi^3/3\right)\right]} = i^n\int^\infty_{-\infty}\text{e}^{is\theta}\frac{\partial^n C(s,\alpha,\beta)}{\partial s^n} \text{d}s,
\end{equation}
where
\begin{equation}
    C(s, \alpha,\beta) = (4\beta)^{-1/3}\exp\left(-is\frac{\alpha}{8\beta}+i\frac{8\beta}{3}\left(\frac{\alpha}{8\beta}\right)^3\right)\text{Ai}(y),
\end{equation}
depends on the Airy function $\text{Ai}(y)$ and its derivative $\text{Ai}'(y)$ where their arguments are given by ${y = (4\beta)^{2/3}[s/4\beta-\left(\alpha/8\beta\right)^2]}$ .
The resulting probability for an electron to emit a photon with energy ratio $x_\omega$ per unit time becomes  \cite{ritus_1985,BERESTETSKII1982,baier1968quasiclassical}
 \begin{multline}
\frac{\text{d}P}{\text{d}x_\omega \text{d}t}(\chi,u)  = - \alpha \frac{m^2}{\epsilon}\left\{ \int_z^\infty \text{Ai}(t)\, \text{d}t \right. \\\left.+ \frac{\text{Ai}'(z)}{z} \left[ 2 + \frac{u^2}{(1+u)}\right]  \right\},
\end{multline}
where $z = [u/\chi]^{2/3}$. To turn this into the LCFA, the asymptotic momentum $p$ is replaced with the (local) classical kinetic momentum, $\pi$, where:
\be 
\pi^{\mu} = p^{\mu}-eA^{\mu} + k^{\mu}(2 eA\cdot p - e^{2}A\cdot A)/(2\,k\cdot p),
\ee 
which can be performed straightforwardly as $p$ only occurs in the constant crossed field rate in the combination $k\cdot p = k\cdot \pi$. In comparison to the LMA, in the LCFA, the discrete sum over harmonics $s$ is replaced by a continuous integral which can be performed analytically. Therefore the LCFA is considerably quicker to evaluate numerically, especially for larger values of $a_{0}$, where many harmonics in the LMA must be included before the sum converges. Also, instead of the electron's instantaneous quasimomentum $q^\mu$ used in the LMA, in the LCFA, only the instantaneous kinetic momentum $p^\mu$, is required. The applicability range of the LCFA has been investigated in the literature, and the condition reported \cite{PhysRevLett.89.094801,DiPiazza2018,BenKing2019}
\be 
\left(\frac{a_{0}^{3}(\phi)u}{\chi(\phi)}\right)^{1/3} \gg 1, \label{eqn:lcfawhen}
\ee
where $a_{0}(\phi)$ and $\chi(\phi)$ are the \emph{local} values of $a_{0}$ and $\chi$.

\subsection{Radiation Integral-BCK model}
This final method provides a general expression to calculate the resulting radiation emitted from a relativistic charged particle moving along a classical trajectory in an arbitrary electromagnetic field \cite{BERESTETSKII1982}. Starting with the semi-classical wave function of a particle moving in an external electromagnetic field 
\begin{equation}
    \Psi = \frac{1}{\sqrt{2\hat{H}}}u(\hat{p})\text{e}^{-i\hat{H}t/\hbar}\phi(\bold{x}),
\end{equation}
where $\hat{H}$ is the Hamiltonian operator. The function $\phi(\bold{x})\propto \text{e}^{-iS/\hbar}$ is the spatial wave function and depends on the classical action $S$. In perturbation theory, the probability to emit a photon with frequency  $\omega$ and wave vector $\bold{k}$ is usually evaluated by squaring the transition matrix element, multiplying by the density of states and summing over the final states of the electron
\begin{equation}
    \text{d}P = \sum_f |M|^2\frac{\text{d}^3\bold{k}}{(2\pi)^3} = \sum_f \left| \int V(t) \text{d}t\right|^2\frac{\text{d}^3\bold{k}}{(2\pi)^3}. \label{eqn:dP1}
\end{equation}
The transition matrix element in \cref{MatrixElement} is the same as in the above expression, but now we separate the integral $\text{d}^4x$ into $\text{d}^3\bold{x}\,\text{d}t$ 
where the element $V(t)$ now becomes a 3-dimensional integral over space,
\begin{equation}
        V(t) = -\frac{e\sqrt{ 2\pi}}{\sqrt{\hbar\omega}}\int {\Psi_{p'}^\dagger(\bold{x},t)}\bold{e}^*\cdot\boldsymbol{\alpha}\text{e}^{i\omega t-i\bold{k\cdot x}}\Psi_p(\bold{x},t)\text{d}^3\bold{x},
\end{equation}
where $\boldsymbol{\alpha}$ is the regular Dirac matrix (see e.g. Ref \cite{BERESTETSKII1982}).
We can write the new matrix element in terms of the operator $\hat{Q}(t)$
\begin{equation}
    \hat{V} = \frac{e\sqrt{ 2\pi}}{\sqrt{\hbar\omega}}\bra{f}\hat{Q}(t)\ket{i}\text{e}^{i\omega t},
\end{equation}
which is defined as
\begin{equation}
    \hat{Q}(t) = \frac{u^\dagger_f(\hat{p}(t))}{\sqrt{2\hat{H}(t)}} \bold{e}^*\cdot\boldsymbol{\alpha}~\text{e}^{-i\bold{k\cdot x}} \frac{u_i(\hat{p}(t))}{\sqrt{2\hat{H}(t)}},
\end{equation}
where factors of the form $\text{e}^{-i\hat{H}t/\hbar}$ have been absorbed, converting the operators into a time-dependent Heisenberg operator. Since the spatial wave functions form a complete basis, we can write ${\sum_f \phi^{\ast}(\bold{x}_2)\phi(\bold{x}_1) = \delta(\bold{x}_2-\bold{x}_1)}$,
which reduces the emission probability to an expectation value given by
\begin{equation}
    \text{d}P = \frac{e^2\text{d}^3\bold{k}}{4\pi^2\hbar\omega}\int \int \bra{i}Q^\dagger(t_2)Q(t_1)\ket{i}\text{e}^{i\omega (t_1-t_2)} \text{d} t_1\text{d} t_2.
\end{equation}
Now the semi-classical operator method developed by Baier and Katkov \cite{baier1968quasiclassical} is employed, where in the product of commuting operators, the operators are replaced by their classical values. Through approximations, the product $\hat{Q^\dagger}\hat{Q}$ is rewritten in terms of commuting operators which are subsequently replaced by their classical non-operator value. The accuracy of this approximation is $O(1/\gamma)$ in a general background, but the approximation is actually exact in a plane wave, making it a powerful tool for simulating relativistic particles colliding with laser pulses, that will be presented in this paper. A detailed derivation of the approximation is shown in Refs. \cite{BERESTETSKII1982,baier1968quasiclassical,Baier}. The result is reduced to a single integral over time (see e.g. \cite{belkacem_1985} and derivations shown in Ref. \cite{kimball_1986} and \cite{TobiasUdvikling}), given by,
\begin{equation}
\label{cue}
\frac{d^2P}{d\hbar\omega d\Omega}=\frac{e^2}{4\pi^3}\left(\frac{\epsilon^{*2}+\epsilon^2}{2\hbar\omega \epsilon^2}\left| \mbf{I} \right|^2 + \frac{\hbar\omega}{2\epsilon^2\gamma^2}\left| J \right|^2\right),
\end{equation}
where $\epsilon^* = \epsilon-\hbar\omega$, (with $\omega^*$ defined via the relation $\epsilon^*\omega^*=\epsilon\omega$) and $I$ and $J$ are given by
\begin{equation}
\label{Iterm}
\mbf{I} = \int_{-\infty}^{\infty} \frac{\bm{n}\times[(\bm{n}-\bm{\beta})\times\dot{\bm{\beta}}]}{(1-\bm{n}\cdot\bm{\beta})^2}e^{i\omega^*(t-\bm{n}\cdot\bm{r})}dt,
\end{equation}
\begin{equation}
\label{Jterm}
J = \int_{-\infty}^{\infty} \frac{\bm{n}\cdot\dot{\bm{\beta}}}{(1-\bm{n}\cdot\bm{\beta})^2}e^{i\omega^*(t-\bm{n}\cdot\bm{r})}dt .
\end{equation}
We call this model the Belkacem, Cue, Kimbal (BCK) model. Here $\bm{n}=(\sin\vartheta\cos\varphi,\sin\vartheta\sin\varphi,\cos\vartheta)$ is the direction of emission with polar and azimuthal angles $\vartheta$ and $\varphi$ and $d\Omega=\sin\vartheta d\vartheta d\varphi$. The classical trajectory is characterized by the instantaneous position $\bm{r}(t)$, the instantaneous velocity $\bm{\beta}(t)$, and the instantaneous acceleration $\dot{\bm{\beta}}(t)$.

As evident from the derivation, no details of the external field or the trajectory has been assumed, only that the trajectory should be described classically, which holds for $1 \gg \chi/\gamma^3$. The Volkov solution to the Dirac equation in a plane-wave background is semi-classical exact, and the BCK method agrees exactly with the QED result in a plane wave. For non-plane-wave backgrounds, from the discussion in Ref.\cite{baier1968quasiclassical}, the angle of the emitted radiation has been assumed to be of the order $1/\gamma$ when determining the commutativity of the operators, and we can therefore expect errors of the order $1/\gamma^2$ when using this method in focused backgrounds. 

When applying the BCK model, it is necessary to determine the trajectory first and then evaluate the spectrum numerically for arbitrary external fields second. (For details of the numerical implementation of the BCK model,  see Ref. \cite{NielsenGPU2020}.) Classically the radiation from an electron can be determined by the Li\'enard-Wiechert radiation integrals \cite{jackson_classical_1998} which is very similar to the BCK model. So the fact that one can define a semi-classical version, which is used to include the nonlinear QED interaction with a plane wave background by integrating over the classical trajectory, might not be too surprising. Lindhard showed \cite{Lind91} that for a spin-0 particle, if the only non-negligible quantum effect is the photon recoil, a simple substitution of the frequency variable in the classical photon number spectrum ($\omega\rightarrow\omega^*$), with $\omega^*\geq0$, regardless of the details of the motion of the particle, reproduced the exact quantum mechanical expression  \cite{Matveev1957}.

When using the BCK model, breaking up the $\Delta t$ integral into smaller sections can be necessary in order to keep the probability of emitting a photon during a section less than one. 
The breaking up of the trajectory into sub-trajectory sections leads to an unphysical contribution in the radiation spectrum due to how the trajectory is patched between the pieces. While this contribution is not present in an electron-laser collision, its origin can be understood on physical grounds. If a particle follows a trajectory which is ballistic between random collisions that each change the momentum by a factor proportional to $|\Delta \boldsymbol{\beta}|$,
 the BCK model can be reduced to \cite{Tobias2019}:
\begin{equation}
    \frac{\text{d}P}{\text{d}\hbar\omega} = \frac{2\alpha}{3\pi\hbar\omega}\gamma^2|\Delta \boldsymbol{\beta}|^2\left(1-\frac{\hbar\omega}{\epsilon}+\frac{3}{4}\frac{\hbar\omega}{\epsilon}\right), \label{eqn:brem1}
\end{equation}
which has the same analytical form as Bethe-Heitler bremsstrahlung formula \cite{PDG_2018}. If one were to use the BCK model to calculate the spectrum from a particle moving through an amorphous material subjected to multiple Coulomb scattering, the overall scaling of the spectrum obtained agrees with what the Bethe-Heitler bremsstrahlung formula \cite{PDG_2018} predicts \cite{Nielsen2020LL}.
This means that instantaneous changes in momentum lead to a bremsstrahlung-like emission spectrum, which will be added on top of any coherence the trajectory might have. 

 If the trajectory becomes too short, the non-physical Bethe-Heitler bremsstrahlung can be significant, which is important to keep in mind when analyzing the resulting spectra, as breaking up the trajectory into smaller pieces is unavoidable because the probability of emitting a photon during a section has to be kept smaller than unity.
 The total spectrum is calculated by adding the probabilities from each small subsection, i.e. the contributions from the subsections are summed incoherently: 
\be
\bigg|\int_{t_{i}}^{t_{f}}  F(t)\,dt\bigg|^{2}
\approx \sum_{n=1}^{N}\bigg|\int_{t^{(n-1)}_{i}}^{t^{(n)}_{f}}  F(t)\, dt\bigg|^{2},
\ee
where $t^{(0)}_{i} = t_{i}$ and $t_{f}^{(N)}=t_{f}$. This means the coherence from the final part of the $(n-1)\trm{th}$ section and the beginning of the $n\trm{th}$ section will be lost. The error introduced in the spectrum is of the order of the contribution on a formation length (described in the following section).

 The Bethe-Heitler effect is absent in the LCFA and LMA rates, because the integration over $\Delta t$ is performed analytically (after an expansion of the probability in $\Delta t$). In high fields (large $a_0$) and long pulses, the laser pulse can no longer be considered a `thin target' and higher orders of nonlinear Compton scattering must be included.

\section{Formation Length}
The formation length, or `coherence time' is the range of values of $\Delta t$ that must be integrated over to effectively include the contribution from the process. A simple approximation for the formation length can be determined by expanding the exponent in emission angle, assuming it to be much smaller than unity (see e.g. \cite{TobiasUdvikling,NielsenGPU2020}) and neglecting terms that vary with the particle position, to give (for a head-on collision):
\begin{equation}\label{FormationQuantum}
    l_f = 2\gamma^2\frac{\epsilon-\hbar \omega}{\epsilon\omega}.
\end{equation}
For our simulations it is important that the trajectory over which one integrates is several formations lengths long.  This is especially the case when applying the BCK model, where a suitable time over which to integrate the trajectory is required.  

From the classical point of view, the formation length can be considered as the distance an electron and the emitted photon travel before they are separated by more than one reduced wavelength. Therefore the formation length (for a head-on collision) can also be expressed as
\begin{equation}
    l_f = 2\gamma^2/\omega. \label{eqn:lfclass1}
\end{equation}
The quantum version of the formation length in \cref{FormationQuantum} can be retrieved from the classical result \cref{eqn:lfclass1} by substituting in $\omega \rightarrow\omega^{*}$, as discussed in the previous section.  A detailed discussion of the concept of the formation lengths is presented in Ref. \cite{UlrikFormation}. 

It should be noted that the formation length can be significantly increased/decreased by contributions at larger values of $\Delta t$ than those captured by the local approximations.
If the particle's trajectory bends considerably outside the $1/\gamma$ light cone, the coherence is broken and the effective formation length is smaller. For the LMA, a similar oscillating phase appears in the derivation (see eq (50) in \cite{TobiasCompton} where we neglect the $\theta^2$ term) which can be written as 
\begin{equation}\label{FormationCompton}
    l_f=\frac{2\gamma^2}{\frac{\epsilon\omega(1+a_0^2/2)}{\epsilon-\hbar \omega}-4\gamma^2 s\omega_l},
\end{equation}
where $\omega_l$ is the frequency of the plane wave background. In this case the position-varying part of the phase is non-negligible and there arises an interesting phenomenon that the formation length diverges exactly when the photon energy is equal to the harmonic frequencies of the nonlinear Compton spectrum. 
This indicates that finite-size effects, such as the finite duration of the pulse, must lead to modifications at these points in the spectrum. (Indeed when the QED probability in a finite plane wave was compared to numerical simulations using the LMA, the error was found to increase at these harmonic edges \cite{Blackburn:2021rqm}.) That the large values of the parameter $a_0$ suppresses the formation length is no surprise, since it is equivalent to the amount of transverse momentum the particle receives during one plane wave cycle. This means that for large $a_0$, particles will be bent outside the light cone breaking the coherence.

\section{Numerical comparison and simulations}
When designing laser-particle experiments, numerical simulations are employed to calculate particle spectra for a range of parameters and in multiple geometries. There is a balance between speed of evaluation and accuracy of simulation of these models. The LCFA is the easiest to employ and the fastest to run but does not capture important spectral features at intermediate intensities $a_{0}\sim O(1)$; the LMA and BCK models are more accurate but if the intensity of the background satisfies  $a_{0}\gtrsim 1$, can take several orders of magnitude longer to evaluate. Therefore, knowing where each model is valid, allows one to switch between models and reduce the evaluation time.

For the LCFA model, the photon spectrum is evaluated at each particle time step based on the local value of $\chi$. In contrast, the LMA divides the trajectory into bigger sections where the average $a_0$ and $\chi$ values are used to evaluate the spectrum for each section. In the LMA, particles can then be pushed consistently according to the ponderomotive force equation \cite{Blackburn:2021cuq,Blackburn:2021rqm}, but here the electrons are modeled as moving through the laser pulse ballistically, which we find is a good approximation for the energies considered. The total LMA spectrum is the sum of the individual spectra from each section. We are working within the `thin target' approximation, where each electron can only undergo single nonlinear Compton scattering.  
As a result, the BCK model can integrate over the entire interference time $\Delta t$. We note that, for all three models, if the intensity and interval of integration are sufficiently large, the calculated probability of emitting a photon can exceed unity. In this case, we can no longer use the thin-target approximation and instead, the trajectories must be broken up into smaller sub-trajectories, which, when integrated over, each give a probability of emission substantially less than unity. This procedure then describes multiple nonlinear Compton scattering.

The BCK model has been applied mainly to strong-field crystal radiation; here we will apply it to electron-laser collisions. First we will benchmark the BCK with the LMA in a monochromatic background, which is exact in this case, to demonstrate the BCK model's reproduction of harmonic structure. Then, we will benchmark with a numerical evaluation of the exact QED expression in a plane wave background; this will confirm the accuracy of the BCK model. Finally, we will compare the models in the collision between an electron and a focused laser background, dispensing with the direct QED evaluation, for which there is currently no solution to a focused background.

\subsection{Monochromatic wave}
\begin{widetext}
\begin{figure}[h!!]
    \centering
    \includegraphics[width = 1.9\linewidth]{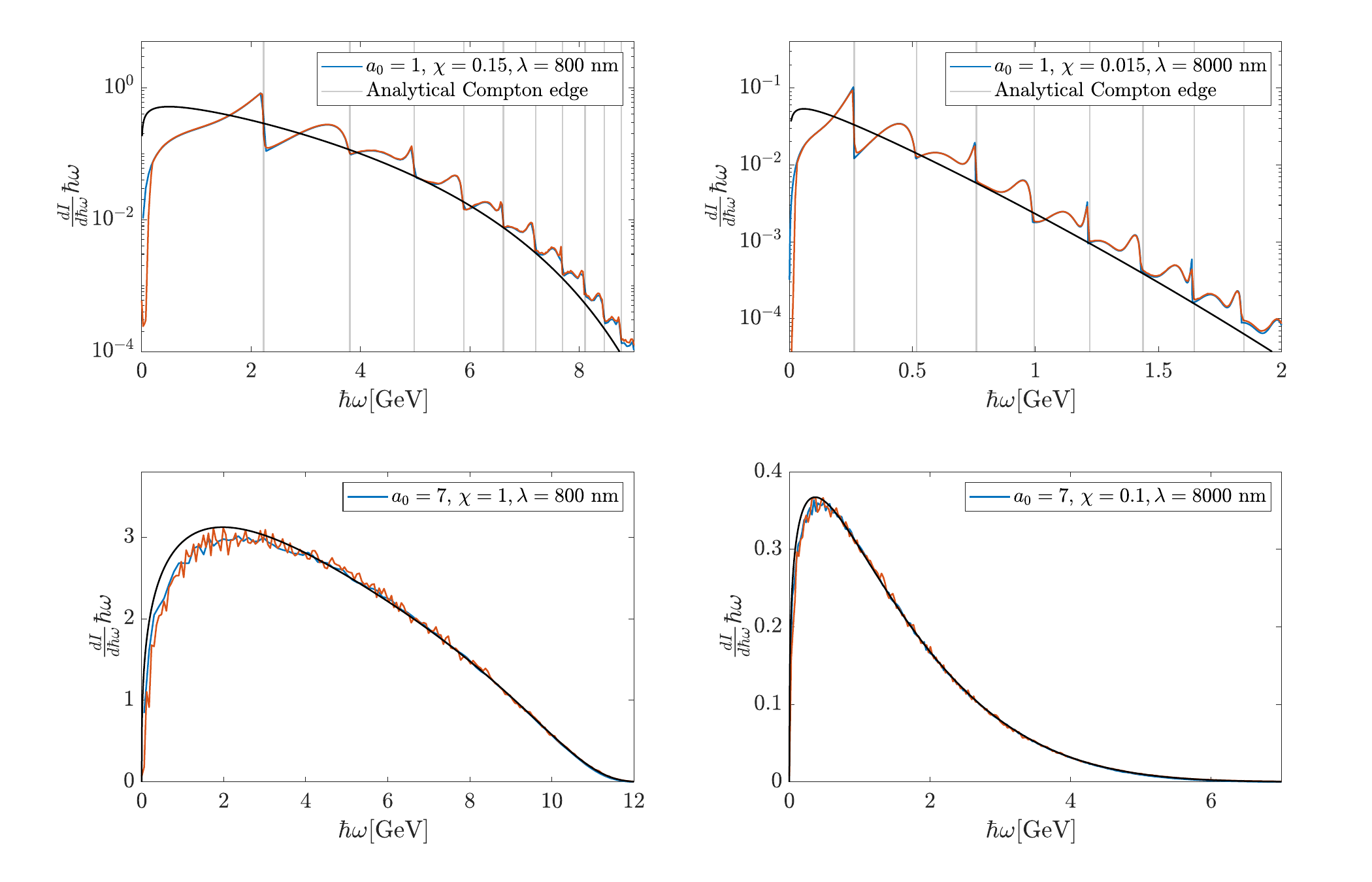}
    \caption{Radiation spectra from a head-on collision between 13 GeV electrons moving 100 fs through a monochromatic wave for different values of $a_0$, $\chi$, and wavelength. The black curve is the radiation spectrum according to the LCFA, the blue curve is the LMA and the red curve is the BCK model. The gridlines mark the position of the harmonics (kinematic edges), as calculated from the formula $\hbar\omega_s = (4s\hbar\omega\epsilon^2)/(m^2(1+a_0^2/2)+4s\hbar\omega\epsilon)$ \cite{E144}}.
    \label{fig:PlaneWaveSim}
\end{figure}
\end{widetext}

In this section, the background is of the form of a linearly polarized monochromatic wave: $\mbf{a}^{\perp} = ma_{0}\{\cos\phi,0\}$, where the vector potential $A = (0,\mbf{a}^{\perp},0)$. In this background, the LMA is exact, therefore the comparison can be used to show the accuracy of the BCK model.  For the comparison, we choose intensity parameters $a_0 = 1$ and $a_0 = 7$ to correspond to the intermediate and high intensity range, respectively, the initial electron energy is 13 GeV 
and the plane-wave wavelength $\lambda$ is 800 or 8000 nm (corresponding to energy parameters $\eta = 0.154$ or $\eta=0.0154$ respectively where $\eta= \hbar k\cdot p/m^{2}$). The simulation is performed for particles moving $100\,\trm{fs}$ through the monochromatic wave. In this case the distance traveled in the field is only important for the BCK model, which integrates the trajectory numerically
, while the LCFA only requires the trajectory to sample the local field strength during one plane wave cycle. In \cref{fig:PlaneWaveSim} we show the radiation spectra from a head on collision between the electron and monochromatic plane wave.

When $a_0 = 1$, we see a confirmation of what would be expected for the intermediate intensity regime, for both wavelengths. The BCK and LMA spectra display harmonic structure, with the $n$th harmonic order corresponding to a net absorption of $n$ laser photons by the electron. The spectrum displays the pattern of sharp harmonics of odd order and more `rounded' harmonics of even order which is typical for a linearly-polarized background (commented on in, e.g. \cite{Titov:2020taw}). In contrast to these models, the LCFA fails to reproduce the harmonic shape, as expected. As already discussed in \cref{FormationCompton}, the formation length diverges at the harmonic edges in the spectrum. This is reflected by the LMA producing sharper harmonic edges, where the BCK model has a smoother structure which reflects harmonic broadening. This can be understood by recalling that the LMA takes into account interference over the period of the fast oscillation, i.e. $\Delta t \approx 2\pi/k^{0}$, but approximates interference over longer periods, whereas in the BCK approach, a finite particle trajectory is integrated numerically and therefore interference on the length scale of the entire trajectory is included. (Sharper edges are produced in plane wave pulses with more cycles.)

For $a_0 = 7$, i.e. in the high-intensity region, there is good agreement between the LMA and BCK spectra where now the strong harmonic peaks appear as small-amplitude, high-frequency noise. Distinguishing these spectra experimentally would be extremely difficult because there is no clear difference even at low energy. 

The comparison at $a_{0}=1$ and wavelength of $\lambda = 8000$ nm reveals an important limitation on the approximate formation length equation \cref{FormationQuantum}.  For these parameters and a 13 GeV electron emitting a 2 GeV photon, the formation length should be $l_f \approx 100$ nm. This would imply that $l_f \ll \lambda$ and therefore harmonic structure in the photon spectrum should not be significant (harmonic structure is due to interference on the length-scale of the wavelength $\lambda$). However, we see that the LMA and BCK indeed have significant harmonic features and therefore there are limitations to using \cref{FormationQuantum} for values of $a_{0} \not \gg 1$. In this case, the formula in \cref{FormationCompton} is required for a more accurate estimate of the formation length.

\begin{figure}[t!]
    \centering
    \includegraphics[width = 0.98\linewidth]{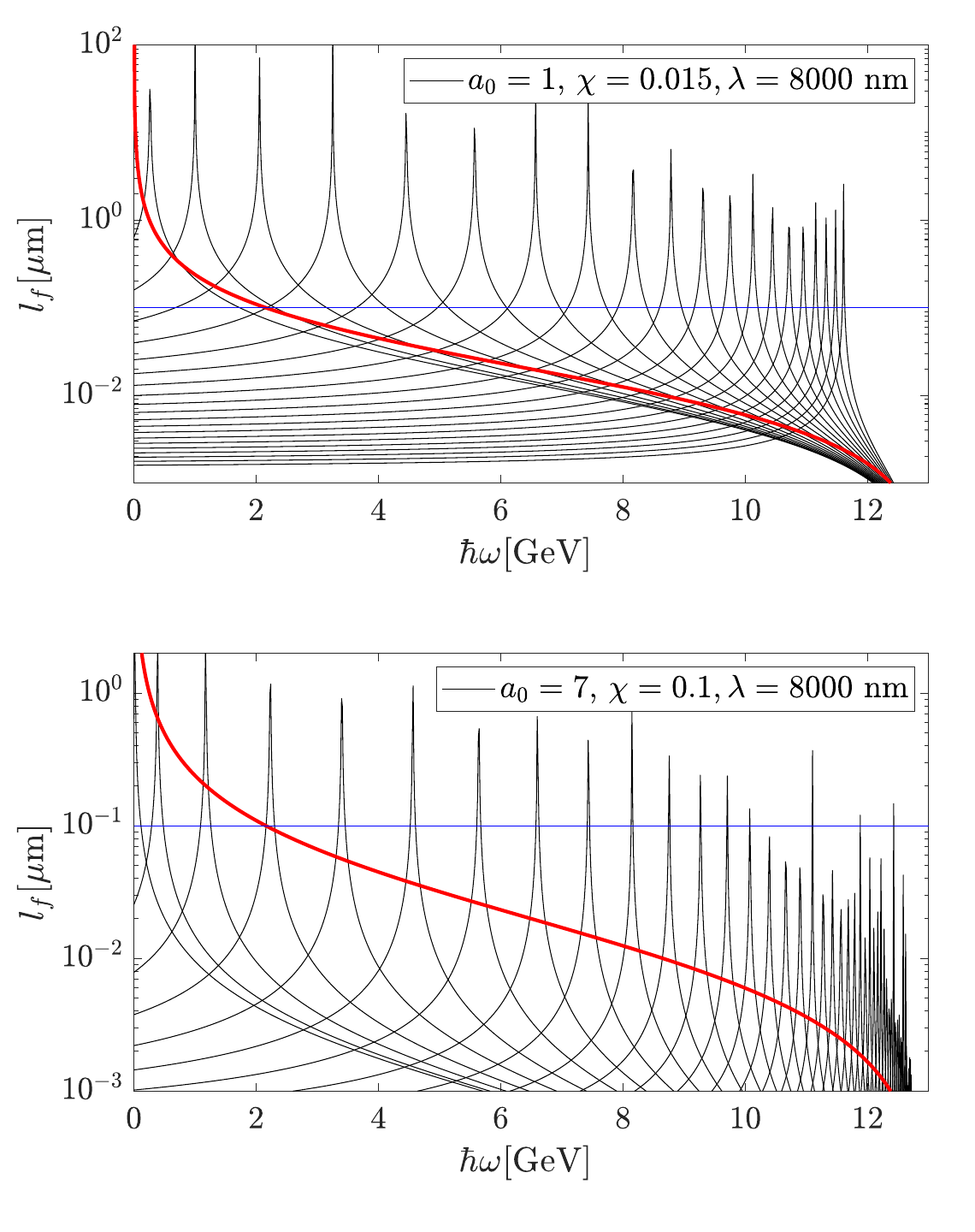}
    \caption{Formation length for a $13$ GeV electron with different values of $a_0$ and $\chi$ moving through a plane wave with 8000 nm wavelength according to \cref{FormationQuantum} (red), and  \cref{FormationCompton} (black) for increasing values of absorbed photons from the external field $s$ throughout the spectrum. The blue line marks a length of $ 0.1$ $\mu$m. The values of $s$ are not equidistant and are chosen to represent the entire spectrum}
    \label{fig:FormationLength}
\end{figure}

In \cref{fig:FormationLength} we show the formation length as a function of emitted photon frequency according to \cref{FormationQuantum,FormationCompton} for the 8000 nm case, $a_0 = 1$ and $a_0 = 7$.
The divergence of the formation length clearly has a large effect in the $a_0 = 1$ case. The diverging peaks, that occur at net number of absorbed of photons $s$ from the external field, span a large region of the spectrum. Increasing the value of $a_0$, increases the transverse momentum a particle receives during a plane wave cycle. If the particle is deflected outside the emission cone, which is of the order $1/\gamma$, the coherence in the motion of the particle is lost. As a result, the effective coherence length becomes smaller than the formation length for larger $a_0$. This $a_0$ suppression of the formation length is clearly visible in the $a_0 = 7$ case, where only a small fraction of the spectrum extends above the region where the field is not constant.

At the low energy end of the spectra, the difference between the models can be understood in terms of the formation length. The formation length becomes much longer for low energy photons for all cases investigated in this paper and each of these models is accurate for different formation length ranges.

\subsection{Finite plane wave pulse}

Before considering the more experimentally-relevant case of a tightly focused laser pulse, we benchmark the models against an evaluation of the exact QED result in a linearly-polarized finite plane-wave pulse of the form:
\be
\mbf{a}^{\perp} =  \left\{\begin{array}{ll}
        ma_{0}\sin^{2}\left(\frac{\vphi}{2N}\right)\left\{\cos\vphi, 0\right\}, &  0\leq \vphi \leq 2\pi N\\
        0 &  \trm{otherwise,}        \end{array}\right.
\ee
 where $N$ is the number of laser cycles. The parameters of the test case are chosen to overlap those at upcoming high-energy experiments. The electron energy is set to  $13\,\trm{GeV}$ and the collision is chosen to be head-on (corresponding to an energy parameter $\eta= 0.154$). The intensity parameter is $a_{0}=1$, which is approximately at the cross-over between the behavior of perturbative and non-perturbative dependency on $a_{0}$. Finally, the two cases of $N=2$  and $N=8$, with full-width-at-half-maximum (FWHM) durations of $2\,\trm{fs}$ and $8\,\trm{fs}$ respectively, are chosen to study short- and long-pulse effects.

\begin{figure}[h!!]
    \centering
    \includegraphics[width=8cm]{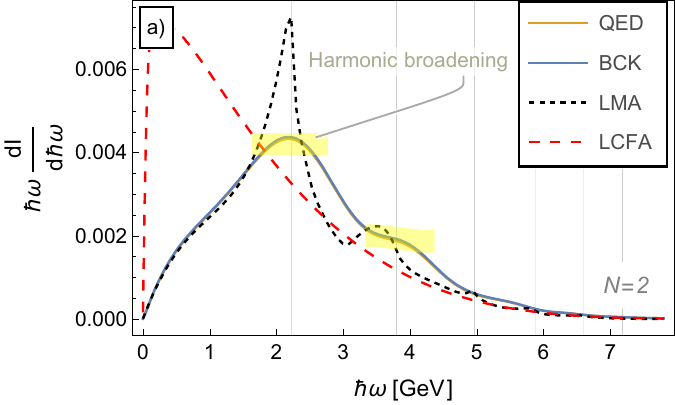}\\\includegraphics[width=8cm]{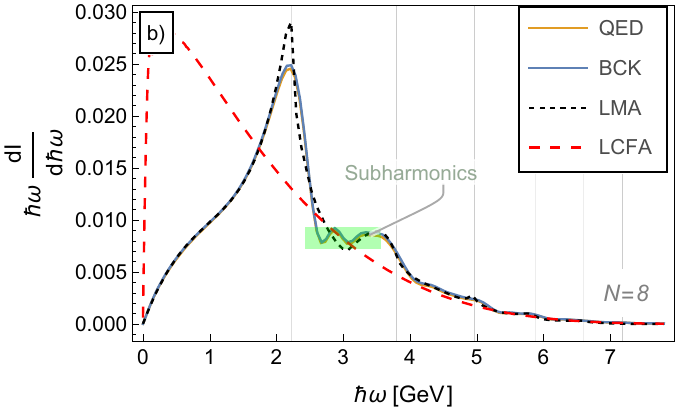}
    \caption{Comparison of direct evaluation of QED expressions, BCK, LMA and LCFA for a head-on collision between an electron with energy $13$ GeV and a sine-squared linearly polarized plane wave with intensity parameter $a_{0}=1$, for: a) $N=2$ cycles and b) $N=8$ cycles.}
    \label{fig:PW1}
\end{figure}

There are a number of conclusions that can be made from the results, illustrated in \cref{fig:PW1}. First, we see reflected in the numerics, the fact that the BCK model is exact in a plane wave; the BCK and QED curves are difficult to distinguish, as one might expect. Second, the BCK model reproduces exactly, the phenomena of harmonic broadening and appearance of subharmonics, in contrast to the LMA, which misses these features. This aspect is straightforward to understand: these effects are to do with the bandwidth of the pulse, and since the shape of the LMA spectrum is independent of the pulse duration (it occurs as a simple prefactor to the rate), it does not contain bandwidth effects. (The LMA has recently been applied to nonlinear Compton in a circularly-polarized background  \cite{Blackburn:2021rqm,King:2020hsk,Fedotov:2022ely}, and the corresponding spectra are similar in nature to the ones presented here.) Third, the error in the LMA is larger for $N=2$ than $N=8$ which is also understandable, since the LMA involves an expansion in the inverse pulse duration and assumes here that $1/N \ll 1$. However, the LMA \emph{does} correctly predict the position of harmonics and well-approximate the overall structure of the spectrum. Finally, the results in \cref{fig:PW1} also demonstrate the well-established \cite{harvey2015testing,DiPiazza2018} fact that the LCFA greatly overestimates the low-energy part of the spectrum.

With the equivalence having been established between the BCK model and the QED result for a plane wave background, we focus on the comparison of the BCK model (which can be applied to non-plane-wave backgrounds), with the LMA and LCFA.

\subsection{Focused laser pulse}
In this section we consider the head-on collision of 13 GeV electrons all with the same initial condition, with a Gaussian focused laser pulse \cite{siegman_lasers_1986,kogelnik_laser_1966} of the form:
\begin{multline}
\label{Paraxial}
\bold{E} = E_0 \frac{w_0}{w(z)}\exp\bigg[-\frac{r^2}{w^2(z)}-2\ln(2)\frac{(t-z)^2}{\tau_0^2}\bigg]\\
\times\Re\bigg\{ \exp\bigg[i\omega(t-z)-\frac{ikr^2}{2R(z)}+i\psi_g(z) \bigg]\bigg\}\textbf{e}_\perp\,.
\end{multline}
Here, $E_0$ is the peak electric field at the focus, $w(z) = w_0\sqrt{1+(z/z_R)^2}$ describes the evolution of the beam width as a function of distance $z$ from the waist, and $z_R = \pi w_0^2/\lambda$ is the Rayleigh length.
Furthermore, the wave-front curvature is described by $R(z) = z[1+(z_R/z)^2]$, $\psi_g(z) = \arctan(z/z_R)$ denotes the Gouy phase, $\Re\{\cdot\}$ denotes the real part of what is inside the braces and 
$\tau_0$ denotes the FWHM pulse duration of the intensity profile.

\begin{widetext}
\begin{figure}[h!!]
    \centering
    \includegraphics[width = 1.9\linewidth]{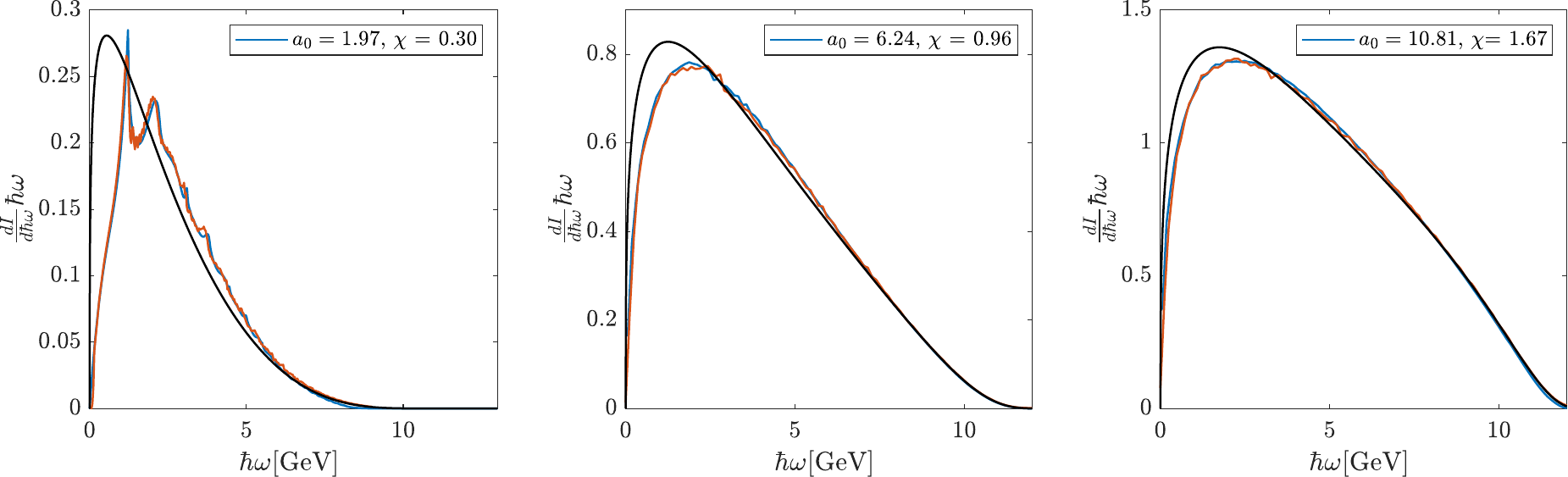}
    \caption{Radiation spectra for 13 GeV electrons colliding head-on with a pulsed laser having a 40 fs FWHM pulse duration and a transverse spot size of $w_0 = 3$ $\mu$m. Legends indicate the maximum $a_0$ and $\chi$ parameter at the focal point of the laser pulse. The black curve is based on the LCFA, the red curve is based on the BCK model and the blue curve is based on the LMA.}
    \label{fig:40fsPulsedSpectrum}
\end{figure}
\end{widetext}

The pulsed laser has a FWHM duration of 40 fs, a transverse spot size of $w_0 = 3\,\mu$m, and a wavelength of 800$\,$nm with intensity parameters: $a_{0}=1.96$, $a_{0}=6.21$ and $a_{0}=10.81$, corresponding to $\chi$ values either side of and close to unity of: $\chi=0.3$, $\chi=0.96$ and $\chi=1.67$ respectively (these values are used as they correspond to a laser pulse energy of $0.1$, $1$ and $3\,\trm{J}$ respectively). The corresponding radiation spectrum is plotted in \cref{fig:40fsPulsedSpectrum}. For $a_{0}=1.96$, harmonics are still visible in the LMA/BCK models. As the laser intensity is increased, the harmonics bunch together and decrease in visibility: for $a_{0}=6.21$ and $a_{0}=10.81$, we see good agreement between all three models in the high-energy tail of the spectrum. However, the discrepancy around the peak of the spectrum is still significant at these intensities, for example in the case $a_0 \approx 11$, this discrepancy persists for energies up to around 3 GeV, i.e. almost $20\%$ of the original electron energy. The disagreement of the low-energy part of the LCFA spectrum with direct evaluation from QED in a plane wave was noted in \cite{DiPiazza2018}; here it is demonstrated for a focused background using the BCK and LMA methods.

\begin{figure}[h!!]
    \centering
    \includegraphics[width = 0.98\linewidth]{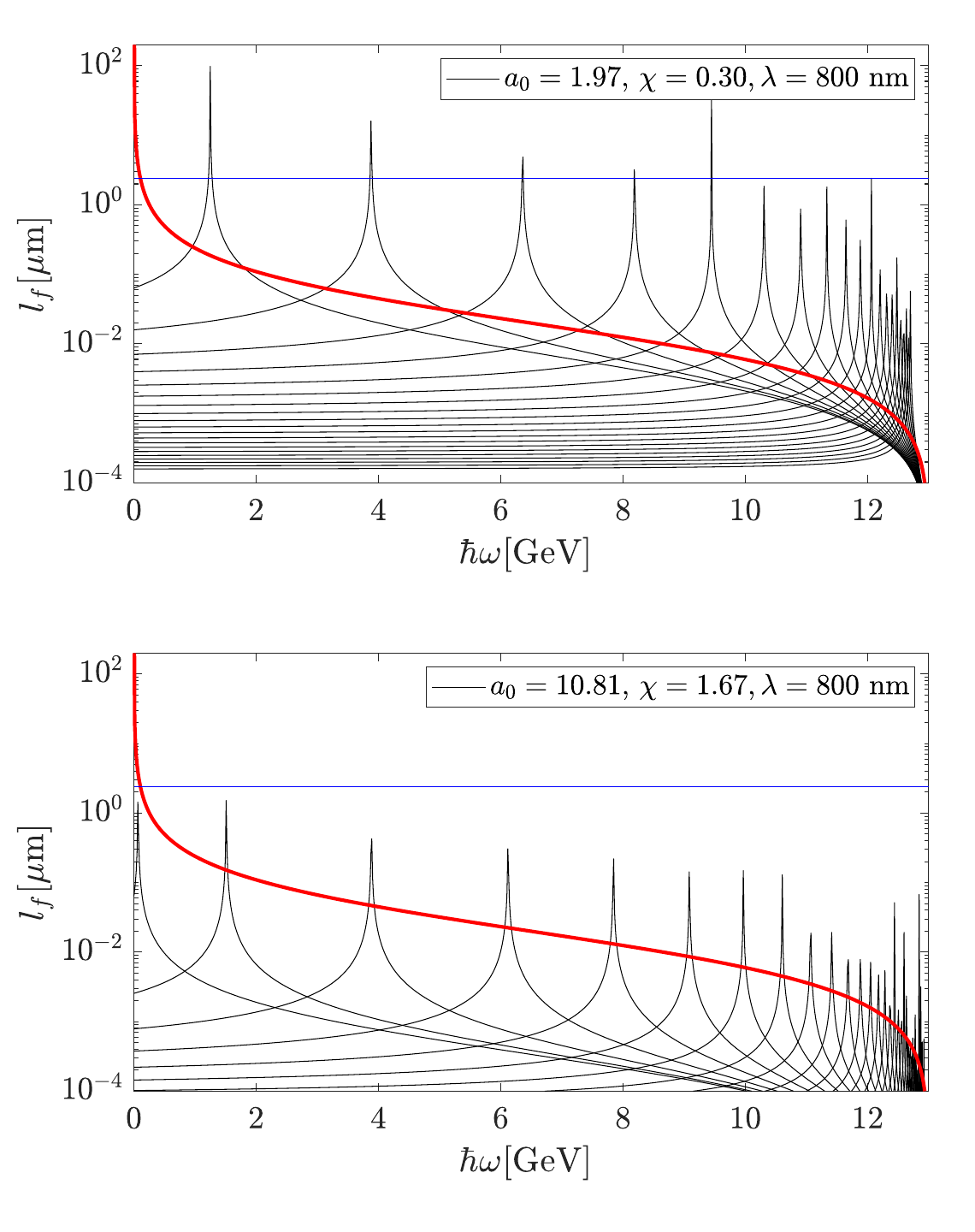}
    \caption{Formation length for a $13$ GeV electron with different values of $a_0$ and $\chi$ moving through a plane wave with 800 nm wavelength according to \cref{FormationQuantum} (red), and according to \cref{FormationCompton} (black) for increasing values of $s$ throughout the spectrum. The blue line marks a length of $ 2.4$ $\mu$m (3 wavelengths). The values of $s$ are not equidistant and are chosen to represent the entire spectrum}
    \label{fig:FormationLengthPulse}
\end{figure}

That the spectra from the BCK and LMA models agree well, can be understood by examining the formation length based on \cref{FormationQuantum,FormationCompton}, which is shown in \cref{fig:FormationLengthPulse}. Since the formation length remains only of the order of the pulse central wavelength, the laser field can be well-approximated by the LMA.

\begin{figure*}[ht]
    \centering
    \includegraphics[width = \linewidth]{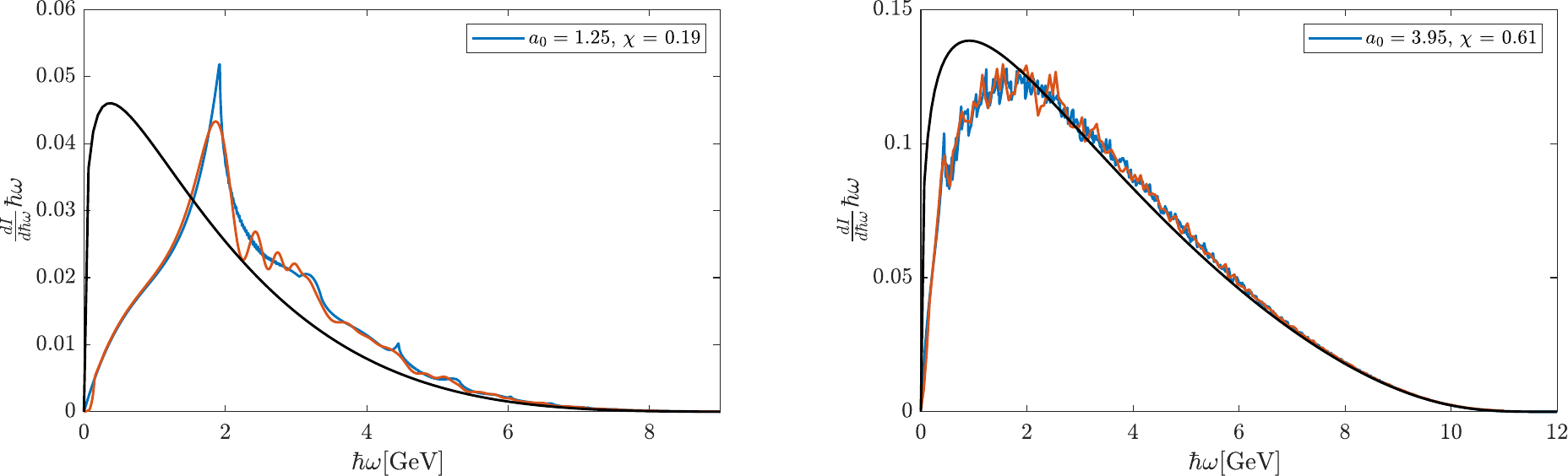}
    \caption{Radiation spectra for 13 GeV electrons colliding head-on with a pulsed laser having a 10 fs FWHM pulse duration and a spot size of $w_0 = 3$ $\mu$m. Legends indicate the maximum $a_0$ and $\chi$ parameter at the focal point of the laser pulse. The black curve is based on the LCFA, the red curve is based on the BCK model and the blue curve is based on the LMA.}
    \label{fig:10fsPulsedSpectrum}
\end{figure*}

With longer laser pulse lengths, the probability of several interactions occurring during a collision between the laser and an electron increases, if an electron has emitted low-energy photons in addition to the measured hard photon
(the contribution of soft and collinear photons to nonlinear Compton scattering has recently been explored in \cite{Edwards:2020npu}). To avoid this situation, we investigate the spectrum produced from a short laser pulse. In \cref{fig:10fsPulsedSpectrum} we show the radiation spectrum from $13$ GeV electrons colliding head on with a pulsed laser having a 10 fs FWHM pulse duration, a transverse spot size of $w_0 = 3$ $\mu$m, and a wavelength of 800 nm for two different laser intensities, $a_{0}=1.25$ and $a_{0}=3.95$, corresponding to $\chi=0.19$ and $\chi=0.61$ respectively. As evident from the $a_0  = 1.25$ case, the difference between the LCFA and the LMA/BCK models has increased as the laser pulse length has decreased. In addition, we now start seeing a difference between the LMA and the BCK models. This is similar to what we observed in the plane-wave benchmarking in \cref{fig:PW1}, and follows from the fact that the LMA can be understood as the leading term in an expansion of the probability in the inverse pulse duration \cite{Torgrimsson:2020gws}; hence for short pulse durations, the error can potentially increase.

For the $a_0 = 3.95$ case, we see the same trend as in previous simulations. The $a_0$ suppression of the formation length ensures that the plane wave amplitude does not change significantly within the reduced formation length, as a consequence the LMA agrees well with the BCK model. (We still see disagreement with the LCFA model starting from the peak moving to lower energy.)

\section{Numerical Computation Time of Models}
\begin{figure}[ht]
    \centering
    \includegraphics[width = \linewidth]{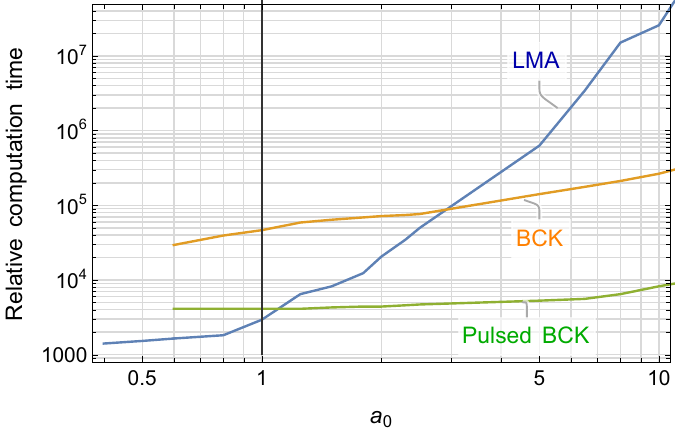}
    \caption{The ratio of the computation time of the LMA and BCK models to the computation time of the LCFA (${1.7\,\times\,10^{-4}}$ s, independent of $a_0$ and $\chi$). The LMA and BCK models were evaluated for 100 photon energies produced by a 13 GeV electron. The LMA and BCK results are for a monochromatic wave, the `pulsed BCK' result is for a 100 fs long trajectory in a Gaussian focused plane wave, with peak intensity $a_{0}$, identical to the parameters used in \cref{fig:40fsPulsedSpectrum}.}
    \label{fig:Timings}
\end{figure}

In this section we compare computation times of the different radiation models described and discuss the parameter space in which each model can or should be used. 
in \cref{fig:Timings} we show the computation time as a function of $a_0$ of 100 photon energies in the spectrum emitted by 13 GeV electrons, in the LMA and BCK models. The computation time is the ratio compared to the LCFA computing time ($10^{-4}$ s), which is independent of $a_0$ and $\chi$. The two cases investigated have parameters identical to what is used in \cref{fig:PlaneWaveSim,fig:40fsPulsedSpectrum} where either a monochromatic plane wave or a Gaussian focused plane wave is used. 

We see that the time it takes to evaluate the LMA scales as a power law with $a_0>1$.  This is because of the increasingly large number of harmonics that must be included before the sum converges. This can be understood physically, because of the `effective mass'  \cite{sengupta52,Brown:1964zzb,Harvey:2012ie}, $m(1+a_{0}^{2}/2)^{1/2}$, the electron acquires in the background. As the intensity increases, the momentum supplied by the background, $sk$, must increase to satisfy energy-momentum conservation, which in turn places a requirement on higher harmonic order, $s$ (and the double Bessel function also takes longer to evaluate for higher harmonic order). The requirement for a higher harmonic order can also be understood mathematically by noting the dependence of the Bessel functions on the argument $z$ in \cref{eqn:z1}. The largest contribution originates from small argument, $z$, which implies $u_{s}\approx u$, or $2s\eta \approx u(1+a_{0}^{2}/2)$, giving a harmonic order $s$ scaling with $\sim a_{0}^{2}$ as $a_{0}$ is increased above $1$.
Because of the exponential scaling with $a_0$, the total computation time will be dominated by the few trajectory grid points that lie at peak $a_0$, and the total computation time will be similar to the monochromatic case, which is why the pulsed plane wave case is left out of \cref{fig:Timings} for the LMA.

The computation time of the BCK model scales linearly with the number of time steps in the trajectory, and by the number of grid points in the numerical integral of the emission angle. For both the monochromatic and pulsed case, the computation time scales linearly with $a_0$. This is because the angular excursion of the particle scales linearly with $a_0$, which in turn increases the area of the emission angle which has to be numerically integrated. 

For a monochromatic plane wave, the particle moves on a trajectory with identical oscillations during each plane wave cycle. The perfect coherence in such a trajectory narrows the angle of emission, and only a few small locations in the angular integral contribute to the spectrum and the resolution of the numerical integration over the emission angle has to be increased. The more plane wave cycles the particle traverses, the narrower these points become, but as discussed earlier, the kinematic Compton edges become sharper as well. In the pulsed case, the value of $a_0$ changes throughout the pulse and the perfect coherence is broken, meaning that fewer points in the angular integral are needed. 

It is evident from this investigation that for cases with moderate and small $a_0$ that vary smoothly, where the harmonic structure is most pronounced the LMA an appropriate model to use, since its computation time in this regime is fairly quick. Since the computation time for the LMA scales disadvantageously for increasing $a_{0}$, but the LCFA becomes more accurate, a hybrid approach using the LMA and LCFA, depending on $a_0$ and $\chi$, becomes advantageous. 
If one requires a very accurate simulation (particularly for lower photon energies) for $a_0 = 1...10$, the BCK is recommended because it scales more favorably with $a_0$ than the LMA. In addition,the BCK should be used for extremely short pulses, as investigated earlier, where envelope effects such as harmonic broadening and subharmonics in the spectrum appear, which are not captured by the LMA.

\section{Conclusion}
We have calculated the photon spectrum from nonlinear Compton scattering in the collision of an electron with a laser pulse. For a plane wave pulse, we demonstrated numerically that the BCK model exactly agrees with the QED result and compared it to predictions using the locally monochromatic approximation (LMA) and the locally constant field approximation (LCFA). We then performed several comparisons of the nonlinear Compton spectrum in the collision of an electron and a focused laser pulse for different intensities, wavelengths and pulse lengths.

From our results, we conclude that the BCK model can be used to calculate high resolution spectra for single nonlinear Compton scattering in a focused laser background. The model includes bandwidth effects such as harmonic broadening and subharmonic structure, which are beyond local approaches such as the LMA and LCFA. The most efficient method of calculating a high resolution spectrum is to use a hybrid approach. Our analysis suggests that the formation length can indicate which model is more appropriate to calculate the spectrum, depending on what part of the field the electron samples as it propagates through a focused laser pulse. Here, we have applied the BCK model to single nonlinear Compton scattering; possible future work includes extending the BCK model to higher orders of emission.

\section*{Acknowledgments}
The authors would like to thank the E144 collaboration and especially Christian Bula and Kostya Shmakov for helping us find and letting us use their nonlinear Compton simulation code, used in the E144 experiment. The numerical results presented in this work were obtained by support from Nvidias GPU grant program. RH acknowledges support from the U.S. National Science Foundation (Grant No. PHY-1535696, and PHY-2012549). BK acknowledges the hospitality of the DESY theory group and support from the Deutsche Forschungsgemeinschaft (DFG, German Research Foundation) under Germany’s Excellence Strategy – EXC 2121 ``Quantum Universe'' – 390833306.

\bibliography{References}

\begin{thebibliography}{84}%
\makeatletter
\providecommand \@ifxundefined [1]{%
 \@ifx{#1\undefined}
}%
\providecommand \@ifnum [1]{%
 \ifnum #1\expandafter \@firstoftwo
 \else \expandafter \@secondoftwo
 \fi
}%
\providecommand \@ifx [1]{%
 \ifx #1\expandafter \@firstoftwo
 \else \expandafter \@secondoftwo
 \fi
}%
\providecommand \natexlab [1]{#1}%
\providecommand \enquote  [1]{``#1''}%
\providecommand \bibnamefont  [1]{#1}%
\providecommand \bibfnamefont [1]{#1}%
\providecommand \citenamefont [1]{#1}%
\providecommand \href@noop [0]{\@secondoftwo}%
\providecommand \href [0]{\begingroup \@sanitize@url \@href}%
\providecommand \@href[1]{\@@startlink{#1}\@@href}%
\providecommand \@@href[1]{\endgroup#1\@@endlink}%
\providecommand \@sanitize@url [0]{\catcode `\\12\catcode `\$12\catcode
  `\&12\catcode `\#12\catcode `\^12\catcode `\_12\catcode `\%12\relax}%
\providecommand \@@startlink[1]{}%
\providecommand \@@endlink[0]{}%
\providecommand \url  [0]{\begingroup\@sanitize@url \@url }%
\providecommand \@url [1]{\endgroup\@href {#1}{\urlprefix }}%
\providecommand \urlprefix  [0]{URL }%
\providecommand \Eprint [0]{\href }%
\providecommand \doibase [0]{http://dx.doi.org/}%
\providecommand \selectlanguage [0]{\@gobble}%
\providecommand \bibinfo  [0]{\@secondoftwo}%
\providecommand \bibfield  [0]{\@secondoftwo}%
\providecommand \translation [1]{[#1]}%
\providecommand \BibitemOpen [0]{}%
\providecommand \bibitemStop [0]{}%
\providecommand \bibitemNoStop [0]{.\EOS\space}%
\providecommand \EOS [0]{\spacefactor3000\relax}%
\providecommand \BibitemShut  [1]{\csname bibitem#1\endcsname}%
\let\auto@bib@innerbib\@empty
\bibitem [{\citenamefont {Burke}\ \emph {et~al.}(1997)\citenamefont {Burke},
  \citenamefont {Field}, \citenamefont {Horton-Smith}, \citenamefont {Spencer},
  \citenamefont {Walz}, \citenamefont {Berridge}, \citenamefont {Bugg},
  \citenamefont {Shmakov}, \citenamefont {Weidemann}, \citenamefont {Bula},
  \citenamefont {McDonald}, \citenamefont {Prebys}, \citenamefont {Bamber},
  \citenamefont {Boege}, \citenamefont {Koffas} \emph
  {et~al.}}]{burke_positron_1997}%
  \BibitemOpen
  \bibfield  {author} {\bibinfo {author} {\bibfnamefont {D.~L.}\ \bibnamefont
  {Burke}}, \bibinfo {author} {\bibfnamefont {R.~C.}\ \bibnamefont {Field}},
  \bibinfo {author} {\bibfnamefont {G.}~\bibnamefont {Horton-Smith}}, \bibinfo
  {author} {\bibfnamefont {J.~E.}\ \bibnamefont {Spencer}}, \bibinfo {author}
  {\bibfnamefont {D.}~\bibnamefont {Walz}}, \bibinfo {author} {\bibfnamefont
  {S.~C.}\ \bibnamefont {Berridge}}, \bibinfo {author} {\bibfnamefont {W.~M.}\
  \bibnamefont {Bugg}}, \bibinfo {author} {\bibfnamefont {K.}~\bibnamefont
  {Shmakov}}, \bibinfo {author} {\bibfnamefont {A.~W.}\ \bibnamefont
  {Weidemann}}, \bibinfo {author} {\bibfnamefont {C.}~\bibnamefont {Bula}},
  \bibinfo {author} {\bibfnamefont {K.~T.}\ \bibnamefont {McDonald}}, \bibinfo
  {author} {\bibfnamefont {E.~J.}\ \bibnamefont {Prebys}}, \bibinfo {author}
  {\bibfnamefont {C.}~\bibnamefont {Bamber}}, \bibinfo {author} {\bibfnamefont
  {S.~J.}\ \bibnamefont {Boege}}, \bibinfo {author} {\bibfnamefont
  {T.}~\bibnamefont {Koffas}},  \emph {et~al.},\ }\href {\doibase
  10.1103/PhysRevLett.79.1626} {\bibfield  {journal} {\bibinfo  {journal}
  {Phys. Rev. Lett.}\ }\textbf {\bibinfo {volume} {79}},\ \bibinfo {pages}
  {1626} (\bibinfo {year} {1997})}\BibitemShut {NoStop}%
\bibitem [{\citenamefont {Bula}\ \emph
  {et~al.}(1996{\natexlab{a}})\citenamefont {Bula}, \citenamefont {McDonald},
  \citenamefont {Prebys}, \citenamefont {Bamber}, \citenamefont {Boege},
  \citenamefont {Kotseroglou}, \citenamefont {Melissinos}, \citenamefont
  {Meyerhofer}, \citenamefont {Ragg}, \citenamefont {Burke}, \citenamefont
  {Field}, \citenamefont {Horton-Smith}, \citenamefont {Odian}, \citenamefont
  {Spencer}, \citenamefont {Walz} \emph {et~al.}}]{bula_observation_1996}%
  \BibitemOpen
  \bibfield  {author} {\bibinfo {author} {\bibfnamefont {C.}~\bibnamefont
  {Bula}}, \bibinfo {author} {\bibfnamefont {K.~T.}\ \bibnamefont {McDonald}},
  \bibinfo {author} {\bibfnamefont {E.~J.}\ \bibnamefont {Prebys}}, \bibinfo
  {author} {\bibfnamefont {C.}~\bibnamefont {Bamber}}, \bibinfo {author}
  {\bibfnamefont {S.}~\bibnamefont {Boege}}, \bibinfo {author} {\bibfnamefont
  {T.}~\bibnamefont {Kotseroglou}}, \bibinfo {author} {\bibfnamefont {A.~C.}\
  \bibnamefont {Melissinos}}, \bibinfo {author} {\bibfnamefont {D.~D.}\
  \bibnamefont {Meyerhofer}}, \bibinfo {author} {\bibfnamefont
  {W.}~\bibnamefont {Ragg}}, \bibinfo {author} {\bibfnamefont {D.~L.}\
  \bibnamefont {Burke}}, \bibinfo {author} {\bibfnamefont {R.~C.}\ \bibnamefont
  {Field}}, \bibinfo {author} {\bibfnamefont {G.}~\bibnamefont {Horton-Smith}},
  \bibinfo {author} {\bibfnamefont {A.~C.}\ \bibnamefont {Odian}}, \bibinfo
  {author} {\bibfnamefont {J.~E.}\ \bibnamefont {Spencer}}, \bibinfo {author}
  {\bibfnamefont {D.}~\bibnamefont {Walz}},  \emph {et~al.},\ }\href {\doibase
  10.1103/PhysRevLett.76.3116} {\bibfield  {journal} {\bibinfo  {journal}
  {Phys. Rev. Lett.}\ }\textbf {\bibinfo {volume} {76}},\ \bibinfo {pages}
  {3116} (\bibinfo {year} {1996}{\natexlab{a}})}\BibitemShut {NoStop}%
\bibitem [{\citenamefont {Poder}\ \emph {et~al.}(2018)\citenamefont {Poder},
  \citenamefont {Tamburini}, \citenamefont {Sarri}, \citenamefont {Di~Piazza},
  \citenamefont {Kuschel}, \citenamefont {Baird}, \citenamefont {Behm},
  \citenamefont {Bohlen}, \citenamefont {Cole}, \citenamefont {Corvan},
  \citenamefont {Duff}, \citenamefont {Gerstmayr}, \citenamefont {Keitel},
  \citenamefont {Krushelnick}, \citenamefont {Mangles} \emph
  {et~al.}}]{poder_experimental_2018}%
  \BibitemOpen
  \bibfield  {author} {\bibinfo {author} {\bibfnamefont {K.}~\bibnamefont
  {Poder}}, \bibinfo {author} {\bibfnamefont {M.}~\bibnamefont {Tamburini}},
  \bibinfo {author} {\bibfnamefont {G.}~\bibnamefont {Sarri}}, \bibinfo
  {author} {\bibfnamefont {A.}~\bibnamefont {Di~Piazza}}, \bibinfo {author}
  {\bibfnamefont {S.}~\bibnamefont {Kuschel}}, \bibinfo {author} {\bibfnamefont
  {C.~D.}\ \bibnamefont {Baird}}, \bibinfo {author} {\bibfnamefont
  {K.}~\bibnamefont {Behm}}, \bibinfo {author} {\bibfnamefont {S.}~\bibnamefont
  {Bohlen}}, \bibinfo {author} {\bibfnamefont {J.~M.}\ \bibnamefont {Cole}},
  \bibinfo {author} {\bibfnamefont {D.~J.}\ \bibnamefont {Corvan}}, \bibinfo
  {author} {\bibfnamefont {M.}~\bibnamefont {Duff}}, \bibinfo {author}
  {\bibfnamefont {E.}~\bibnamefont {Gerstmayr}}, \bibinfo {author}
  {\bibfnamefont {C.~H.}\ \bibnamefont {Keitel}}, \bibinfo {author}
  {\bibfnamefont {K.}~\bibnamefont {Krushelnick}}, \bibinfo {author}
  {\bibfnamefont {S.~P.~D.}\ \bibnamefont {Mangles}},  \emph {et~al.},\ }\href
  {\doibase 10.1103/PhysRevX.8.031004} {\bibfield  {journal} {\bibinfo
  {journal} {Phys. Rev. X}\ }\textbf {\bibinfo {volume} {8}},\ \bibinfo {pages}
  {031004} (\bibinfo {year} {2018})}\BibitemShut {NoStop}%
\bibitem [{\citenamefont {Cole}\ \emph {et~al.}(2018)\citenamefont {Cole},
  \citenamefont {Behm}, \citenamefont {Gerstmayr}, \citenamefont {Blackburn},
  \citenamefont {Wood}, \citenamefont {Baird}, \citenamefont {Duff},
  \citenamefont {Harvey}, \citenamefont {Ilderton}, \citenamefont {Joglekar},
  \citenamefont {Krushelnick}, \citenamefont {Kuschel}, \citenamefont
  {Marklund}, \citenamefont {McKenna}, \citenamefont {Murphy} \emph
  {et~al.}}]{cole_experimental_2018}%
  \BibitemOpen
  \bibfield  {author} {\bibinfo {author} {\bibfnamefont {J.~M.}\ \bibnamefont
  {Cole}}, \bibinfo {author} {\bibfnamefont {K.~T.}\ \bibnamefont {Behm}},
  \bibinfo {author} {\bibfnamefont {E.}~\bibnamefont {Gerstmayr}}, \bibinfo
  {author} {\bibfnamefont {T.~G.}\ \bibnamefont {Blackburn}}, \bibinfo {author}
  {\bibfnamefont {J.~C.}\ \bibnamefont {Wood}}, \bibinfo {author}
  {\bibfnamefont {C.~D.}\ \bibnamefont {Baird}}, \bibinfo {author}
  {\bibfnamefont {M.~J.}\ \bibnamefont {Duff}}, \bibinfo {author}
  {\bibfnamefont {C.}~\bibnamefont {Harvey}}, \bibinfo {author} {\bibfnamefont
  {A.}~\bibnamefont {Ilderton}}, \bibinfo {author} {\bibfnamefont {A.~S.}\
  \bibnamefont {Joglekar}}, \bibinfo {author} {\bibfnamefont {K.}~\bibnamefont
  {Krushelnick}}, \bibinfo {author} {\bibfnamefont {S.}~\bibnamefont
  {Kuschel}}, \bibinfo {author} {\bibfnamefont {M.}~\bibnamefont {Marklund}},
  \bibinfo {author} {\bibfnamefont {P.}~\bibnamefont {McKenna}}, \bibinfo
  {author} {\bibfnamefont {C.~D.}\ \bibnamefont {Murphy}},  \emph {et~al.},\
  }\href {\doibase 10.1103/PhysRevX.8.011020} {\bibfield  {journal} {\bibinfo
  {journal} {Phys. Rev. X}\ }\textbf {\bibinfo {volume} {8}},\ \bibinfo {pages}
  {011020} (\bibinfo {year} {2018})}\BibitemShut {NoStop}%
\bibitem [{\citenamefont {Gonsalves}\ \emph {et~al.}(2019)\citenamefont
  {Gonsalves}, \citenamefont {Nakamura}, \citenamefont {Daniels}, \citenamefont
  {Benedetti}, \citenamefont {Pieronek}, \citenamefont {de~Raadt},
  \citenamefont {Steinke}, \citenamefont {Bin}, \citenamefont {Bulanov},
  \citenamefont {van Tilborg}, \citenamefont {Geddes}, \citenamefont
  {Schroeder}, \citenamefont {T\'oth}, \citenamefont {Esarey}, \citenamefont
  {Swanson}, \citenamefont {Fan-Chiang}, \citenamefont {Bagdasarov},
  \citenamefont {Bobrova}, \citenamefont {Gasilov}, \citenamefont {Korn},
  \citenamefont {Sasorov},\ and\ \citenamefont
  {Leemans}}]{gonsalves_pw_8gev_2019}%
  \BibitemOpen
  \bibfield  {author} {\bibinfo {author} {\bibfnamefont {A.~J.}\ \bibnamefont
  {Gonsalves}}, \bibinfo {author} {\bibfnamefont {K.}~\bibnamefont {Nakamura}},
  \bibinfo {author} {\bibfnamefont {J.}~\bibnamefont {Daniels}}, \bibinfo
  {author} {\bibfnamefont {C.}~\bibnamefont {Benedetti}}, \bibinfo {author}
  {\bibfnamefont {C.}~\bibnamefont {Pieronek}}, \bibinfo {author}
  {\bibfnamefont {T.~C.~H.}\ \bibnamefont {de~Raadt}}, \bibinfo {author}
  {\bibfnamefont {S.}~\bibnamefont {Steinke}}, \bibinfo {author} {\bibfnamefont
  {J.~H.}\ \bibnamefont {Bin}}, \bibinfo {author} {\bibfnamefont {S.~S.}\
  \bibnamefont {Bulanov}}, \bibinfo {author} {\bibfnamefont {J.}~\bibnamefont
  {van Tilborg}}, \bibinfo {author} {\bibfnamefont {C.~G.~R.}\ \bibnamefont
  {Geddes}}, \bibinfo {author} {\bibfnamefont {C.~B.}\ \bibnamefont
  {Schroeder}}, \bibinfo {author} {\bibfnamefont {C.}~\bibnamefont {T\'oth}},
  \bibinfo {author} {\bibfnamefont {E.}~\bibnamefont {Esarey}}, \bibinfo
  {author} {\bibfnamefont {K.}~\bibnamefont {Swanson}}, \bibinfo {author}
  {\bibfnamefont {L.}~\bibnamefont {Fan-Chiang}}, \bibinfo {author}
  {\bibfnamefont {G.}~\bibnamefont {Bagdasarov}}, \bibinfo {author}
  {\bibfnamefont {N.}~\bibnamefont {Bobrova}}, \bibinfo {author} {\bibfnamefont
  {V.}~\bibnamefont {Gasilov}}, \bibinfo {author} {\bibfnamefont
  {G.}~\bibnamefont {Korn}}, \bibinfo {author} {\bibfnamefont {P.}~\bibnamefont
  {Sasorov}}, \ and\ \bibinfo {author} {\bibfnamefont {W.~P.}\ \bibnamefont
  {Leemans}},\ }\href {\doibase 10.1103/PhysRevLett.122.084801} {\bibfield
  {journal} {\bibinfo  {journal} {Phys. Rev. Lett.}\ }\textbf {\bibinfo
  {volume} {122}},\ \bibinfo {pages} {084801} (\bibinfo {year}
  {2019})}\BibitemShut {NoStop}%
\bibitem [{\citenamefont {Leemans}\ \emph {et~al.}(2014)\citenamefont
  {Leemans}, \citenamefont {Gonsalves}, \citenamefont {{H.-S. Mao}},
  \citenamefont {Nakamura}, \citenamefont {Benedetti}, \citenamefont
  {Schroeder}, \citenamefont {T{\'{o}}th}, \citenamefont {Daniels},
  \citenamefont {Mittelberger}, \citenamefont {Bulanov}, \citenamefont {{J.-L.
  Vay}}, \citenamefont {Geddes},\ and\ \citenamefont
  {Esarey}}]{leemans_multi-gev_2014}%
  \BibitemOpen
  \bibfield  {author} {\bibinfo {author} {\bibfnamefont {W.~P.}\ \bibnamefont
  {Leemans}}, \bibinfo {author} {\bibfnamefont {A.~J.}\ \bibnamefont
  {Gonsalves}}, \bibinfo {author} {\bibnamefont {{H.-S. Mao}}}, \bibinfo
  {author} {\bibfnamefont {K.}~\bibnamefont {Nakamura}}, \bibinfo {author}
  {\bibfnamefont {C.}~\bibnamefont {Benedetti}}, \bibinfo {author}
  {\bibfnamefont {C.~B.}\ \bibnamefont {Schroeder}}, \bibinfo {author}
  {\bibfnamefont {C.}~\bibnamefont {T{\'{o}}th}}, \bibinfo {author}
  {\bibfnamefont {J.}~\bibnamefont {Daniels}}, \bibinfo {author} {\bibfnamefont
  {D.~E.}\ \bibnamefont {Mittelberger}}, \bibinfo {author} {\bibfnamefont
  {S.~S.}\ \bibnamefont {Bulanov}}, \bibinfo {author} {\bibnamefont {{J.-L.
  Vay}}}, \bibinfo {author} {\bibfnamefont {C.~G.~R.}\ \bibnamefont {Geddes}},
  \ and\ \bibinfo {author} {\bibfnamefont {E.}~\bibnamefont {Esarey}},\ }\href
  {\doibase 10.1103/PhysRevLett.113.245002} {\bibfield  {journal} {\bibinfo
  {journal} {Phys. Rev. Lett.}\ }\textbf {\bibinfo {volume} {113}},\ \bibinfo
  {pages} {245002} (\bibinfo {year} {2014})}\BibitemShut {NoStop}%
\bibitem [{\citenamefont {Wang}\ \emph {et~al.}(2013)\citenamefont {Wang},
  \citenamefont {Zgadzaj}, \citenamefont {Fazel}, \citenamefont {Li},
  \citenamefont {Yi}, \citenamefont {Zhang}, \citenamefont {Henderson},
  \citenamefont {Chang}, \citenamefont {Korzekwa}, \citenamefont {Tsai} \emph
  {et~al.}}]{wang2013quasi}%
  \BibitemOpen
  \bibfield  {author} {\bibinfo {author} {\bibfnamefont {X.}~\bibnamefont
  {Wang}}, \bibinfo {author} {\bibfnamefont {R.}~\bibnamefont {Zgadzaj}},
  \bibinfo {author} {\bibfnamefont {N.}~\bibnamefont {Fazel}}, \bibinfo
  {author} {\bibfnamefont {Z.}~\bibnamefont {Li}}, \bibinfo {author}
  {\bibfnamefont {S.}~\bibnamefont {Yi}}, \bibinfo {author} {\bibfnamefont
  {X.}~\bibnamefont {Zhang}}, \bibinfo {author} {\bibfnamefont
  {W.}~\bibnamefont {Henderson}}, \bibinfo {author} {\bibfnamefont {Y.-Y.}\
  \bibnamefont {Chang}}, \bibinfo {author} {\bibfnamefont {R.}~\bibnamefont
  {Korzekwa}}, \bibinfo {author} {\bibfnamefont {H.-E.}\ \bibnamefont {Tsai}},
  \emph {et~al.},\ }\href {\doibase 10.1038/ncomms2988} {\bibfield  {journal}
  {\bibinfo  {journal} {Nature communications}\ }\textbf {\bibinfo {volume}
  {4}},\ \bibinfo {pages} {1} (\bibinfo {year} {2013})}\BibitemShut {NoStop}%
\bibitem [{\citenamefont {Kim}\ \emph {et~al.}(2013)\citenamefont {Kim},
  \citenamefont {Pae}, \citenamefont {Cha}, \citenamefont {Kim}, \citenamefont
  {Yu}, \citenamefont {Sung}, \citenamefont {Lee}, \citenamefont {Jeong},\ and\
  \citenamefont {Lee}}]{kim2013enhancement}%
  \BibitemOpen
  \bibfield  {author} {\bibinfo {author} {\bibfnamefont {H.~T.}\ \bibnamefont
  {Kim}}, \bibinfo {author} {\bibfnamefont {K.~H.}\ \bibnamefont {Pae}},
  \bibinfo {author} {\bibfnamefont {H.~J.}\ \bibnamefont {Cha}}, \bibinfo
  {author} {\bibfnamefont {I.~J.}\ \bibnamefont {Kim}}, \bibinfo {author}
  {\bibfnamefont {T.~J.}\ \bibnamefont {Yu}}, \bibinfo {author} {\bibfnamefont
  {J.~H.}\ \bibnamefont {Sung}}, \bibinfo {author} {\bibfnamefont {S.~K.}\
  \bibnamefont {Lee}}, \bibinfo {author} {\bibfnamefont {T.~M.}\ \bibnamefont
  {Jeong}}, \ and\ \bibinfo {author} {\bibfnamefont {J.}~\bibnamefont {Lee}},\
  }\href@noop {} {\bibfield  {journal} {\bibinfo  {journal} {Physical review
  letters}\ }\textbf {\bibinfo {volume} {111}},\ \bibinfo {pages} {165002}
  (\bibinfo {year} {2013})}\BibitemShut {NoStop}%
\bibitem [{\citenamefont {Tanaka}\ \emph {et~al.}(2020)\citenamefont {Tanaka},
  \citenamefont {Spohr}, \citenamefont {Balabanski}, \citenamefont {Balascuta},
  \citenamefont {Capponi}, \citenamefont {Cernaianu}, \citenamefont {Cuciuc},
  \citenamefont {Cucoanes}, \citenamefont {Dancus}, \citenamefont {Dhal},
  \citenamefont {Diaconescu}, \citenamefont {Doria}, \citenamefont {Ghenuche},
  \citenamefont {Ghita}, \citenamefont {Kisyov}, \citenamefont {Nastasa},
  \citenamefont {Ong}, \citenamefont {Rotaru}, \citenamefont {Sangwan},
  \citenamefont {Söderström}, \citenamefont {Stutman}, \citenamefont
  {Suliman}, \citenamefont {Tesileanu}, \citenamefont {Tudor}, \citenamefont
  {Tsoneva}, \citenamefont {Ur}, \citenamefont {Ursescu},\ and\ \citenamefont
  {Zamfir}}]{tanaka_current_2020}%
  \BibitemOpen
  \bibfield  {author} {\bibinfo {author} {\bibfnamefont {K.~A.}\ \bibnamefont
  {Tanaka}}, \bibinfo {author} {\bibfnamefont {K.~M.}\ \bibnamefont {Spohr}},
  \bibinfo {author} {\bibfnamefont {D.~L.}\ \bibnamefont {Balabanski}},
  \bibinfo {author} {\bibfnamefont {S.}~\bibnamefont {Balascuta}}, \bibinfo
  {author} {\bibfnamefont {L.}~\bibnamefont {Capponi}}, \bibinfo {author}
  {\bibfnamefont {M.~O.}\ \bibnamefont {Cernaianu}}, \bibinfo {author}
  {\bibfnamefont {M.}~\bibnamefont {Cuciuc}}, \bibinfo {author} {\bibfnamefont
  {A.}~\bibnamefont {Cucoanes}}, \bibinfo {author} {\bibfnamefont
  {I.}~\bibnamefont {Dancus}}, \bibinfo {author} {\bibfnamefont
  {A.}~\bibnamefont {Dhal}}, \bibinfo {author} {\bibfnamefont {B.}~\bibnamefont
  {Diaconescu}}, \bibinfo {author} {\bibfnamefont {D.}~\bibnamefont {Doria}},
  \bibinfo {author} {\bibfnamefont {P.}~\bibnamefont {Ghenuche}}, \bibinfo
  {author} {\bibfnamefont {D.~G.}\ \bibnamefont {Ghita}}, \bibinfo {author}
  {\bibfnamefont {S.}~\bibnamefont {Kisyov}}, \bibinfo {author} {\bibfnamefont
  {V.}~\bibnamefont {Nastasa}}, \bibinfo {author} {\bibfnamefont {J.~F.}\
  \bibnamefont {Ong}}, \bibinfo {author} {\bibfnamefont {F.}~\bibnamefont
  {Rotaru}}, \bibinfo {author} {\bibfnamefont {D.}~\bibnamefont {Sangwan}},
  \bibinfo {author} {\bibfnamefont {P.-A.}\ \bibnamefont {Söderström}},
  \bibinfo {author} {\bibfnamefont {D.}~\bibnamefont {Stutman}}, \bibinfo
  {author} {\bibfnamefont {G.}~\bibnamefont {Suliman}}, \bibinfo {author}
  {\bibfnamefont {O.}~\bibnamefont {Tesileanu}}, \bibinfo {author}
  {\bibfnamefont {L.}~\bibnamefont {Tudor}}, \bibinfo {author} {\bibfnamefont
  {N.}~\bibnamefont {Tsoneva}}, \bibinfo {author} {\bibfnamefont {C.~A.}\
  \bibnamefont {Ur}}, \bibinfo {author} {\bibfnamefont {D.}~\bibnamefont
  {Ursescu}}, \ and\ \bibinfo {author} {\bibfnamefont {N.~V.}\ \bibnamefont
  {Zamfir}},\ }\href {\doibase 10.1063/1.5093535} {\bibfield  {journal}
  {\bibinfo  {journal} {Matter and Radiation at Extremes}\ }\textbf {\bibinfo
  {volume} {5}},\ \bibinfo {pages} {024402} (\bibinfo {year}
  {2020})}\BibitemShut {NoStop}%
\bibitem [{\citenamefont {Turcu}\ \emph {et~al.}(2015)\citenamefont {Turcu},
  \citenamefont {Balascuta}, \citenamefont {Negoita}, \citenamefont
  {Jaroszynski},\ and\ \citenamefont {McKenna}}]{turcu_strong_2015}%
  \BibitemOpen
  \bibfield  {author} {\bibinfo {author} {\bibfnamefont {I.~C.~E.}\
  \bibnamefont {Turcu}}, \bibinfo {author} {\bibfnamefont {S.}~\bibnamefont
  {Balascuta}}, \bibinfo {author} {\bibfnamefont {F.}~\bibnamefont {Negoita}},
  \bibinfo {author} {\bibfnamefont {D.}~\bibnamefont {Jaroszynski}}, \ and\
  \bibinfo {author} {\bibfnamefont {P.}~\bibnamefont {McKenna}},\ }\href
  {\doibase 10.1063/1.4909613} {\bibfield  {journal} {\bibinfo  {journal} {AIP
  Conference Proceedings}\ }\textbf {\bibinfo {volume} {1645}},\ \bibinfo
  {pages} {416} (\bibinfo {year} {2015})}\BibitemShut {NoStop}%
\bibitem [{\citenamefont {Maksimchuk}\ \emph {et~al.}(2020)\citenamefont
  {Maksimchuk}, \citenamefont {Nees}, \citenamefont {Kalinchenko},
  \citenamefont {Hou}, \citenamefont {Ma}, \citenamefont {McKelvey},
  \citenamefont {Shi}, \citenamefont {Jovanovic}, \citenamefont {Kuranz},
  \citenamefont {Thomas} \emph {et~al.}}]{maksimchuk2020zeus}%
  \BibitemOpen
  \bibfield  {author} {\bibinfo {author} {\bibfnamefont {A.}~\bibnamefont
  {Maksimchuk}}, \bibinfo {author} {\bibfnamefont {J.}~\bibnamefont {Nees}},
  \bibinfo {author} {\bibfnamefont {G.}~\bibnamefont {Kalinchenko}}, \bibinfo
  {author} {\bibfnamefont {B.}~\bibnamefont {Hou}}, \bibinfo {author}
  {\bibfnamefont {Y.}~\bibnamefont {Ma}}, \bibinfo {author} {\bibfnamefont
  {A.}~\bibnamefont {McKelvey}}, \bibinfo {author} {\bibfnamefont
  {T.}~\bibnamefont {Shi}}, \bibinfo {author} {\bibfnamefont {I.}~\bibnamefont
  {Jovanovic}}, \bibinfo {author} {\bibfnamefont {C.}~\bibnamefont {Kuranz}},
  \bibinfo {author} {\bibfnamefont {A.}~\bibnamefont {Thomas}},  \emph
  {et~al.},\ }\href@noop {} {\bibfield  {journal} {\bibinfo  {journal}
  {Bulletin of the American Physical Society}\ } (\bibinfo {year}
  {2020})}\BibitemShut {NoStop}%
\bibitem [{\citenamefont {Cartlidge}(2018)}]{cartlidge_light_2018}%
  \BibitemOpen
  \bibfield  {author} {\bibinfo {author} {\bibfnamefont {E.}~\bibnamefont
  {Cartlidge}},\ }\href {\doibase 10.1126/science.359.6374.382} {\bibfield
  {journal} {\bibinfo  {journal} {Science}\ }\textbf {\bibinfo {volume}
  {359}},\ \bibinfo {pages} {382} (\bibinfo {year} {2018})}\BibitemShut
  {NoStop}%
\bibitem [{\citenamefont {Weber}\ \emph {et~al.}(2017)\citenamefont {Weber},
  \citenamefont {Bechet}, \citenamefont {Borneis}, \citenamefont {Brabec},
  \citenamefont {Bučka}, \citenamefont {Chacon-Golcher}, \citenamefont
  {Ciappina}, \citenamefont {DeMarco}, \citenamefont {Fajstavr}, \citenamefont
  {Falk}, \citenamefont {Garcia}, \citenamefont {Grosz}, \citenamefont {Gu},
  \citenamefont {Hernandez}, \citenamefont {Holec}, \citenamefont {Janečka},
  \citenamefont {Jantač}, \citenamefont {Jirka}, \citenamefont {Kadlecova},
  \citenamefont {Khikhlukha}, \citenamefont {Klimo}, \citenamefont {Korn},
  \citenamefont {Kramer}, \citenamefont {Kumar}, \citenamefont {Lastovička},
  \citenamefont {Lutoslawski}, \citenamefont {Morejon}, \citenamefont
  {Olšovcová}, \citenamefont {Rajdl}, \citenamefont {Renner}, \citenamefont
  {Rus}, \citenamefont {Singh}, \citenamefont {Šmid}, \citenamefont {Sokol},
  \citenamefont {Versaci}, \citenamefont {Vrána}, \citenamefont {Vranic},
  \citenamefont {Vyskočil}, \citenamefont {Wolf},\ and\ \citenamefont
  {Yu}}]{weber_p3_2017}%
  \BibitemOpen
  \bibfield  {author} {\bibinfo {author} {\bibfnamefont {S.}~\bibnamefont
  {Weber}}, \bibinfo {author} {\bibfnamefont {S.}~\bibnamefont {Bechet}},
  \bibinfo {author} {\bibfnamefont {S.}~\bibnamefont {Borneis}}, \bibinfo
  {author} {\bibfnamefont {L.}~\bibnamefont {Brabec}}, \bibinfo {author}
  {\bibfnamefont {M.}~\bibnamefont {Bučka}}, \bibinfo {author} {\bibfnamefont
  {E.}~\bibnamefont {Chacon-Golcher}}, \bibinfo {author} {\bibfnamefont
  {M.}~\bibnamefont {Ciappina}}, \bibinfo {author} {\bibfnamefont
  {M.}~\bibnamefont {DeMarco}}, \bibinfo {author} {\bibfnamefont
  {A.}~\bibnamefont {Fajstavr}}, \bibinfo {author} {\bibfnamefont
  {K.}~\bibnamefont {Falk}}, \bibinfo {author} {\bibfnamefont {E.-R.}\
  \bibnamefont {Garcia}}, \bibinfo {author} {\bibfnamefont {J.}~\bibnamefont
  {Grosz}}, \bibinfo {author} {\bibfnamefont {Y.-J.}\ \bibnamefont {Gu}},
  \bibinfo {author} {\bibfnamefont {J.-C.}\ \bibnamefont {Hernandez}}, \bibinfo
  {author} {\bibfnamefont {M.}~\bibnamefont {Holec}}, \bibinfo {author}
  {\bibfnamefont {P.}~\bibnamefont {Janečka}}, \bibinfo {author}
  {\bibfnamefont {M.}~\bibnamefont {Jantač}}, \bibinfo {author} {\bibfnamefont
  {M.}~\bibnamefont {Jirka}}, \bibinfo {author} {\bibfnamefont
  {H.}~\bibnamefont {Kadlecova}}, \bibinfo {author} {\bibfnamefont
  {D.}~\bibnamefont {Khikhlukha}}, \bibinfo {author} {\bibfnamefont
  {O.}~\bibnamefont {Klimo}}, \bibinfo {author} {\bibfnamefont
  {G.}~\bibnamefont {Korn}}, \bibinfo {author} {\bibfnamefont {D.}~\bibnamefont
  {Kramer}}, \bibinfo {author} {\bibfnamefont {D.}~\bibnamefont {Kumar}},
  \bibinfo {author} {\bibfnamefont {T.}~\bibnamefont {Lastovička}}, \bibinfo
  {author} {\bibfnamefont {P.}~\bibnamefont {Lutoslawski}}, \bibinfo {author}
  {\bibfnamefont {L.}~\bibnamefont {Morejon}}, \bibinfo {author} {\bibfnamefont
  {V.}~\bibnamefont {Olšovcová}}, \bibinfo {author} {\bibfnamefont
  {M.}~\bibnamefont {Rajdl}}, \bibinfo {author} {\bibfnamefont
  {O.}~\bibnamefont {Renner}}, \bibinfo {author} {\bibfnamefont
  {B.}~\bibnamefont {Rus}}, \bibinfo {author} {\bibfnamefont {S.}~\bibnamefont
  {Singh}}, \bibinfo {author} {\bibfnamefont {M.}~\bibnamefont {Šmid}},
  \bibinfo {author} {\bibfnamefont {M.}~\bibnamefont {Sokol}}, \bibinfo
  {author} {\bibfnamefont {R.}~\bibnamefont {Versaci}}, \bibinfo {author}
  {\bibfnamefont {R.}~\bibnamefont {Vrána}}, \bibinfo {author} {\bibfnamefont
  {M.}~\bibnamefont {Vranic}}, \bibinfo {author} {\bibfnamefont
  {J.}~\bibnamefont {Vyskočil}}, \bibinfo {author} {\bibfnamefont
  {A.}~\bibnamefont {Wolf}}, \ and\ \bibinfo {author} {\bibfnamefont
  {Q.}~\bibnamefont {Yu}},\ }\href {\doibase 10.1016/j.mre.2017.03.003}
  {\bibfield  {journal} {\bibinfo  {journal} {Matter and Radiation at
  Extremes}\ }\textbf {\bibinfo {volume} {2}},\ \bibinfo {pages} {149}
  (\bibinfo {year} {2017})}\BibitemShut {NoStop}%
\bibitem [{\citenamefont {Leemans}\ \emph {et~al.}(2010)\citenamefont
  {Leemans}, \citenamefont {Duarte}, \citenamefont {Esarey}, \citenamefont
  {Fournier}, \citenamefont {Geddes}, \citenamefont {Lockhart}, \citenamefont
  {Schroeder}, \citenamefont {T{\'o}th}, \citenamefont {Vay},\ and\
  \citenamefont {Zimmermann}}]{leemans2010berkeley}%
  \BibitemOpen
  \bibfield  {author} {\bibinfo {author} {\bibfnamefont {W.}~\bibnamefont
  {Leemans}}, \bibinfo {author} {\bibfnamefont {R.}~\bibnamefont {Duarte}},
  \bibinfo {author} {\bibfnamefont {E.}~\bibnamefont {Esarey}}, \bibinfo
  {author} {\bibfnamefont {S.}~\bibnamefont {Fournier}}, \bibinfo {author}
  {\bibfnamefont {C.}~\bibnamefont {Geddes}}, \bibinfo {author} {\bibfnamefont
  {D.}~\bibnamefont {Lockhart}}, \bibinfo {author} {\bibfnamefont
  {C.}~\bibnamefont {Schroeder}}, \bibinfo {author} {\bibfnamefont
  {C.}~\bibnamefont {T{\'o}th}}, \bibinfo {author} {\bibfnamefont {J.-L.}\
  \bibnamefont {Vay}}, \ and\ \bibinfo {author} {\bibfnamefont
  {S.}~\bibnamefont {Zimmermann}},\ }in\ \href@noop {} {\emph {\bibinfo
  {booktitle} {AIP Conference Proceedings}}},\ Vol.\ \bibinfo {volume} {1299}\
  (\bibinfo {organization} {American Institute of Physics},\ \bibinfo {year}
  {2010})\ pp.\ \bibinfo {pages} {3--11}\BibitemShut {NoStop}%
\bibitem [{\citenamefont {Furry}(1951)}]{Furry:1951zz}%
  \BibitemOpen
  \bibfield  {author} {\bibinfo {author} {\bibfnamefont {W.~H.}\ \bibnamefont
  {Furry}},\ }\href {\doibase 10.1103/PhysRev.81.915} {\bibfield  {journal}
  {\bibinfo  {journal} {Phys. Rev.}\ }\textbf {\bibinfo {volume} {81}},\
  \bibinfo {pages} {115} (\bibinfo {year} {1951})}\BibitemShut {NoStop}%
\bibitem [{\citenamefont {Ritus}(1985)}]{ritus_1985}%
  \BibitemOpen
  \bibfield  {author} {\bibinfo {author} {\bibfnamefont {V.~I.}\ \bibnamefont
  {Ritus}},\ }\href {\doibase 10.1007/BF01120220} {\bibfield  {journal}
  {\bibinfo  {journal} {J. Sov. Laser Res.}\ }\textbf {\bibinfo {volume} {6}},\
  \bibinfo {pages} {497} (\bibinfo {year} {1985})}\BibitemShut {NoStop}%
\bibitem [{\citenamefont {{Ehlotzky}}\ \emph {et~al.}(2009)\citenamefont
  {{Ehlotzky}}, \citenamefont {{Krajewska}},\ and\ \citenamefont
  {{Kami{\'n}ski}}}]{2009RPPh...72d6401E}%
  \BibitemOpen
  \bibfield  {author} {\bibinfo {author} {\bibfnamefont {F.}~\bibnamefont
  {{Ehlotzky}}}, \bibinfo {author} {\bibfnamefont {K.}~\bibnamefont
  {{Krajewska}}}, \ and\ \bibinfo {author} {\bibfnamefont {J.~Z.}\ \bibnamefont
  {{Kami{\'n}ski}}},\ }\href {\doibase 10.1088/0034-4885/72/4/046401}
  {\bibfield  {journal} {\bibinfo  {journal} {Reports on Progress in Physics}\
  }\textbf {\bibinfo {volume} {72}},\ \bibinfo {eid} {046401} (\bibinfo {year}
  {2009})}\BibitemShut {NoStop}%
\bibitem [{\citenamefont {Di~Piazza}\ \emph {et~al.}(2012)\citenamefont
  {Di~Piazza}, \citenamefont {M\"uller}, \citenamefont {Hatsagortsyan},\ and\
  \citenamefont {Keitel}}]{AntoninoReview}%
  \BibitemOpen
  \bibfield  {author} {\bibinfo {author} {\bibfnamefont {A.}~\bibnamefont
  {Di~Piazza}}, \bibinfo {author} {\bibfnamefont {C.}~\bibnamefont {M\"uller}},
  \bibinfo {author} {\bibfnamefont {K.~Z.}\ \bibnamefont {Hatsagortsyan}}, \
  and\ \bibinfo {author} {\bibfnamefont {C.~H.}\ \bibnamefont {Keitel}},\
  }\href {\doibase 10.1103/RevModPhys.84.1177} {\bibfield  {journal} {\bibinfo
  {journal} {Rev. Mod. Phys.}\ }\textbf {\bibinfo {volume} {84}},\ \bibinfo
  {pages} {1177} (\bibinfo {year} {2012})}\BibitemShut {NoStop}%
\bibitem [{\citenamefont {Narozhny}\ and\ \citenamefont
  {Fedotov}(2015)}]{Narozhny:2015vsb}%
  \BibitemOpen
  \bibfield  {author} {\bibinfo {author} {\bibfnamefont {N.~B.}\ \bibnamefont
  {Narozhny}}\ and\ \bibinfo {author} {\bibfnamefont {A.~M.}\ \bibnamefont
  {Fedotov}},\ }\href {\doibase 10.1080/00107514.2015.1028768} {\bibfield
  {journal} {\bibinfo  {journal} {Contemp. Phys.}\ }\textbf {\bibinfo {volume}
  {56}},\ \bibinfo {pages} {249} (\bibinfo {year} {2015})}\BibitemShut
  {NoStop}%
\bibitem [{\citenamefont {Fedotov}\ \emph {et~al.}(2022)\citenamefont
  {Fedotov}, \citenamefont {Ilderton}, \citenamefont {Karbstein}, \citenamefont
  {King}, \citenamefont {Seipt}, \citenamefont {Taya},\ and\ \citenamefont
  {Torgrimsson}}]{Fedotov:2022ely}%
  \BibitemOpen
  \bibfield  {author} {\bibinfo {author} {\bibfnamefont {A.}~\bibnamefont
  {Fedotov}}, \bibinfo {author} {\bibfnamefont {A.}~\bibnamefont {Ilderton}},
  \bibinfo {author} {\bibfnamefont {F.}~\bibnamefont {Karbstein}}, \bibinfo
  {author} {\bibfnamefont {B.}~\bibnamefont {King}}, \bibinfo {author}
  {\bibfnamefont {D.}~\bibnamefont {Seipt}}, \bibinfo {author} {\bibfnamefont
  {H.}~\bibnamefont {Taya}}, \ and\ \bibinfo {author} {\bibfnamefont
  {G.}~\bibnamefont {Torgrimsson}},\ }\href@noop {} {\  (\bibinfo {year}
  {2022})},\ \Eprint {http://arxiv.org/abs/2203.00019} {arXiv:2203.00019
  [hep-ph]} \BibitemShut {NoStop}%
\bibitem [{\citenamefont {Elkina}\ \emph {et~al.}(2011)\citenamefont {Elkina},
  \citenamefont {Fedotov}, \citenamefont {Kostyukov}, \citenamefont {Legkov},
  \citenamefont {Narozhny}, \citenamefont {Nerush},\ and\ \citenamefont
  {Ruhl}}]{Elkina:2010up}%
  \BibitemOpen
  \bibfield  {author} {\bibinfo {author} {\bibfnamefont {N.~V.}\ \bibnamefont
  {Elkina}}, \bibinfo {author} {\bibfnamefont {A.~M.}\ \bibnamefont {Fedotov}},
  \bibinfo {author} {\bibfnamefont {I.~Y.}\ \bibnamefont {Kostyukov}}, \bibinfo
  {author} {\bibfnamefont {M.~V.}\ \bibnamefont {Legkov}}, \bibinfo {author}
  {\bibfnamefont {N.~B.}\ \bibnamefont {Narozhny}}, \bibinfo {author}
  {\bibfnamefont {E.~N.}\ \bibnamefont {Nerush}}, \ and\ \bibinfo {author}
  {\bibfnamefont {H.}~\bibnamefont {Ruhl}},\ }\href {\doibase
  10.1103/PhysRevSTAB.14.054401} {\bibfield  {journal} {\bibinfo  {journal}
  {Phys. Rev. ST Accel. Beams}\ }\textbf {\bibinfo {volume} {14}},\ \bibinfo
  {pages} {054401} (\bibinfo {year} {2011})},\ \Eprint
  {http://arxiv.org/abs/1010.4528} {arXiv:1010.4528 [hep-ph]} \BibitemShut
  {NoStop}%
\bibitem [{\citenamefont {Gonoskov}\ \emph {et~al.}(2015)\citenamefont
  {Gonoskov}, \citenamefont {Bastrakov}, \citenamefont {Efimenko},
  \citenamefont {Ilderton}, \citenamefont {Marklund}, \citenamefont {Meyerov},
  \citenamefont {Muraviev}, \citenamefont {Sergeev}, \citenamefont {Surmin},\
  and\ \citenamefont {Wallin}}]{Gonoskov:2014mda}%
  \BibitemOpen
  \bibfield  {author} {\bibinfo {author} {\bibfnamefont {A.}~\bibnamefont
  {Gonoskov}}, \bibinfo {author} {\bibfnamefont {S.}~\bibnamefont {Bastrakov}},
  \bibinfo {author} {\bibfnamefont {E.}~\bibnamefont {Efimenko}}, \bibinfo
  {author} {\bibfnamefont {A.}~\bibnamefont {Ilderton}}, \bibinfo {author}
  {\bibfnamefont {M.}~\bibnamefont {Marklund}}, \bibinfo {author}
  {\bibfnamefont {I.}~\bibnamefont {Meyerov}}, \bibinfo {author} {\bibfnamefont
  {A.}~\bibnamefont {Muraviev}}, \bibinfo {author} {\bibfnamefont
  {A.}~\bibnamefont {Sergeev}}, \bibinfo {author} {\bibfnamefont
  {I.}~\bibnamefont {Surmin}}, \ and\ \bibinfo {author} {\bibfnamefont
  {E.}~\bibnamefont {Wallin}},\ }\href {\doibase 10.1103/PhysRevE.92.023305}
  {\bibfield  {journal} {\bibinfo  {journal} {Phys. Rev. E}\ }\textbf {\bibinfo
  {volume} {92}},\ \bibinfo {pages} {023305} (\bibinfo {year} {2015})},\
  \Eprint {http://arxiv.org/abs/1412.6426} {arXiv:1412.6426 [physics.plasm-ph]}
  \BibitemShut {NoStop}%
\bibitem [{\citenamefont {Ridgers}\ \emph {et~al.}(2014)\citenamefont
  {Ridgers}, \citenamefont {Kirk}, \citenamefont {Duclous}, \citenamefont
  {Blackburn}, \citenamefont {Brady}, \citenamefont {Bennett}, \citenamefont
  {Arber},\ and\ \citenamefont {Bell}}]{RIDGERS2014273}%
  \BibitemOpen
  \bibfield  {author} {\bibinfo {author} {\bibfnamefont {C.}~\bibnamefont
  {Ridgers}}, \bibinfo {author} {\bibfnamefont {J.}~\bibnamefont {Kirk}},
  \bibinfo {author} {\bibfnamefont {R.}~\bibnamefont {Duclous}}, \bibinfo
  {author} {\bibfnamefont {T.}~\bibnamefont {Blackburn}}, \bibinfo {author}
  {\bibfnamefont {C.}~\bibnamefont {Brady}}, \bibinfo {author} {\bibfnamefont
  {K.}~\bibnamefont {Bennett}}, \bibinfo {author} {\bibfnamefont
  {T.}~\bibnamefont {Arber}}, \ and\ \bibinfo {author} {\bibfnamefont
  {A.}~\bibnamefont {Bell}},\ }\href {\doibase
  https://doi.org/10.1016/j.jcp.2013.12.007} {\bibfield  {journal} {\bibinfo
  {journal} {Journal of Computational Physics}\ }\textbf {\bibinfo {volume}
  {260}},\ \bibinfo {pages} {273} (\bibinfo {year} {2014})}\BibitemShut
  {NoStop}%
\bibitem [{\citenamefont {Lobet}\ \emph {et~al.}(2016)\citenamefont {Lobet},
  \citenamefont {d{\textquotesingle}Humi{\`{e}}res}, \citenamefont {Grech},
  \citenamefont {Ruyer}, \citenamefont {Davoine},\ and\ \citenamefont
  {Gremillet}}]{Lobet2016}%
  \BibitemOpen
  \bibfield  {author} {\bibinfo {author} {\bibfnamefont {M.}~\bibnamefont
  {Lobet}}, \bibinfo {author} {\bibfnamefont {E.}~\bibnamefont
  {d{\textquotesingle}Humi{\`{e}}res}}, \bibinfo {author} {\bibfnamefont
  {M.}~\bibnamefont {Grech}}, \bibinfo {author} {\bibfnamefont
  {C.}~\bibnamefont {Ruyer}}, \bibinfo {author} {\bibfnamefont
  {X.}~\bibnamefont {Davoine}}, \ and\ \bibinfo {author} {\bibfnamefont
  {L.}~\bibnamefont {Gremillet}},\ }\href {\doibase
  10.1088/1742-6596/688/1/012058} {\bibfield  {journal} {\bibinfo  {journal}
  {Journal of Physics: Conference Series}\ }\textbf {\bibinfo {volume} {688}},\
  \bibinfo {pages} {012058} (\bibinfo {year} {2016})}\BibitemShut {NoStop}%
\bibitem [{\citenamefont {Gonoskov}\ \emph {et~al.}(2021)\citenamefont
  {Gonoskov}, \citenamefont {Blackburn}, \citenamefont {Marklund},\ and\
  \citenamefont {Bulanov}}]{Gonoskov:2021hwf}%
  \BibitemOpen
  \bibfield  {author} {\bibinfo {author} {\bibfnamefont {A.}~\bibnamefont
  {Gonoskov}}, \bibinfo {author} {\bibfnamefont {T.~G.}\ \bibnamefont
  {Blackburn}}, \bibinfo {author} {\bibfnamefont {M.}~\bibnamefont {Marklund}},
  \ and\ \bibinfo {author} {\bibfnamefont {S.~S.}\ \bibnamefont {Bulanov}},\
  }\href@noop {} {\  (\bibinfo {year} {2021})},\ \Eprint
  {http://arxiv.org/abs/2107.02161} {arXiv:2107.02161 [physics.plasm-ph]}
  \BibitemShut {NoStop}%
\bibitem [{\citenamefont {Harvey}\ \emph
  {et~al.}(2015{\natexlab{a}})\citenamefont {Harvey}, \citenamefont
  {Ilderton},\ and\ \citenamefont {King}}]{BenKing2015}%
  \BibitemOpen
  \bibfield  {author} {\bibinfo {author} {\bibfnamefont {C.~N.}\ \bibnamefont
  {Harvey}}, \bibinfo {author} {\bibfnamefont {A.}~\bibnamefont {Ilderton}}, \
  and\ \bibinfo {author} {\bibfnamefont {B.}~\bibnamefont {King}},\ }\href
  {\doibase 10.1103/PhysRevA.91.013822} {\bibfield  {journal} {\bibinfo
  {journal} {Phys. Rev. A}\ }\textbf {\bibinfo {volume} {91}},\ \bibinfo
  {pages} {013822} (\bibinfo {year} {2015}{\natexlab{a}})}\BibitemShut
  {NoStop}%
\bibitem [{\citenamefont {Di~Piazza}\ \emph {et~al.}(2018)\citenamefont
  {Di~Piazza}, \citenamefont {Tamburini}, \citenamefont {Meuren},\ and\
  \citenamefont {Keitel}}]{DiPiazza2018}%
  \BibitemOpen
  \bibfield  {author} {\bibinfo {author} {\bibfnamefont {A.}~\bibnamefont
  {Di~Piazza}}, \bibinfo {author} {\bibfnamefont {M.}~\bibnamefont
  {Tamburini}}, \bibinfo {author} {\bibfnamefont {S.}~\bibnamefont {Meuren}}, \
  and\ \bibinfo {author} {\bibfnamefont {C.~H.}\ \bibnamefont {Keitel}},\
  }\href {\doibase 10.1103/PhysRevA.98.012134} {\bibfield  {journal} {\bibinfo
  {journal} {Phys. Rev. A}\ }\textbf {\bibinfo {volume} {98}},\ \bibinfo
  {pages} {012134} (\bibinfo {year} {2018})}\BibitemShut {NoStop}%
\bibitem [{\citenamefont {Aleksandrov}\ \emph {et~al.}(2019)\citenamefont
  {Aleksandrov}, \citenamefont {Plunien},\ and\ \citenamefont
  {Shabaev}}]{Aleksandrov:2018zso}%
  \BibitemOpen
  \bibfield  {author} {\bibinfo {author} {\bibfnamefont {I.~A.}\ \bibnamefont
  {Aleksandrov}}, \bibinfo {author} {\bibfnamefont {G.}~\bibnamefont
  {Plunien}}, \ and\ \bibinfo {author} {\bibfnamefont {V.~M.}\ \bibnamefont
  {Shabaev}},\ }\href {\doibase 10.1103/PhysRevD.99.016020} {\bibfield
  {journal} {\bibinfo  {journal} {Phys. Rev. D}\ }\textbf {\bibinfo {volume}
  {99}},\ \bibinfo {pages} {016020} (\bibinfo {year} {2019})},\ \Eprint
  {http://arxiv.org/abs/1811.01419} {arXiv:1811.01419 [hep-ph]} \BibitemShut
  {NoStop}%
\bibitem [{\citenamefont {Di~Piazza}\ \emph {et~al.}(2019)\citenamefont
  {Di~Piazza}, \citenamefont {Tamburini}, \citenamefont {Meuren},\ and\
  \citenamefont {Keitel}}]{DiPiazza2019}%
  \BibitemOpen
  \bibfield  {author} {\bibinfo {author} {\bibfnamefont {A.}~\bibnamefont
  {Di~Piazza}}, \bibinfo {author} {\bibfnamefont {M.}~\bibnamefont
  {Tamburini}}, \bibinfo {author} {\bibfnamefont {S.}~\bibnamefont {Meuren}}, \
  and\ \bibinfo {author} {\bibfnamefont {C.~H.}\ \bibnamefont {Keitel}},\
  }\href {\doibase 10.1103/PhysRevA.99.022125} {\bibfield  {journal} {\bibinfo
  {journal} {Phys. Rev. A}\ }\textbf {\bibinfo {volume} {99}},\ \bibinfo
  {pages} {022125} (\bibinfo {year} {2019})}\BibitemShut {NoStop}%
\bibitem [{\citenamefont {Ilderton}\ \emph {et~al.}(2019)\citenamefont
  {Ilderton}, \citenamefont {King},\ and\ \citenamefont {Seipt}}]{BenKing2019}%
  \BibitemOpen
  \bibfield  {author} {\bibinfo {author} {\bibfnamefont {A.}~\bibnamefont
  {Ilderton}}, \bibinfo {author} {\bibfnamefont {B.}~\bibnamefont {King}}, \
  and\ \bibinfo {author} {\bibfnamefont {D.}~\bibnamefont {Seipt}},\ }\href
  {\doibase 10.1103/PhysRevA.99.042121} {\bibfield  {journal} {\bibinfo
  {journal} {Phys. Rev. A}\ }\textbf {\bibinfo {volume} {99}},\ \bibinfo
  {pages} {042121} (\bibinfo {year} {2019})}\BibitemShut {NoStop}%
\bibitem [{\citenamefont {Raicher}\ \emph {et~al.}(2021)\citenamefont
  {Raicher}, \citenamefont {Lv}, \citenamefont {Keitel},\ and\ \citenamefont
  {Hatsagortsyan}}]{Raicher:2020nkq}%
  \BibitemOpen
  \bibfield  {author} {\bibinfo {author} {\bibfnamefont {E.}~\bibnamefont
  {Raicher}}, \bibinfo {author} {\bibfnamefont {Q.~Z.}\ \bibnamefont {Lv}},
  \bibinfo {author} {\bibfnamefont {C.~H.}\ \bibnamefont {Keitel}}, \ and\
  \bibinfo {author} {\bibfnamefont {K.~Z.}\ \bibnamefont {Hatsagortsyan}},\
  }\href {\doibase 10.1103/PhysRevResearch.3.013214} {\bibfield  {journal}
  {\bibinfo  {journal} {Phys. Rev. Res.}\ }\textbf {\bibinfo {volume} {3}},\
  \bibinfo {pages} {013214} (\bibinfo {year} {2021})},\ \Eprint
  {http://arxiv.org/abs/2003.06217} {arXiv:2003.06217 [physics.plasm-ph]}
  \BibitemShut {NoStop}%
\bibitem [{\citenamefont {Torgrimsson}(2021)}]{Torgrimsson:2020gws}%
  \BibitemOpen
  \bibfield  {author} {\bibinfo {author} {\bibfnamefont {G.}~\bibnamefont
  {Torgrimsson}},\ }\href {\doibase 10.1088/1367-2630/abf274} {\bibfield
  {journal} {\bibinfo  {journal} {New J. Phys.}\ }\textbf {\bibinfo {volume}
  {23}},\ \bibinfo {pages} {065001} (\bibinfo {year} {2021})},\ \Eprint
  {http://arxiv.org/abs/2012.12701} {arXiv:2012.12701 [hep-ph]} \BibitemShut
  {NoStop}%
\bibitem [{\citenamefont {Meuren}(2019)}]{E320Talk}%
  \BibitemOpen
  \bibfield  {author} {\bibinfo {author} {\bibfnamefont {S.}~\bibnamefont
  {Meuren}},\ }\href
  {https://conf.slac.stanford.edu/facet-2-2019/sites/facet-2-2019.conf.slac.stanford.edu/files/basic-page-docs/sfqed_2019.pdf}
  {\enquote {\bibinfo {title} {Probing strong-field qed at facet-ii (slac
  e-320)},}\ } (\bibinfo {year} {2019}),\ \bibinfo {note} {talk}\BibitemShut
  {NoStop}%
\bibitem [{\citenamefont {Abramowicz}\ \emph {et~al.}(2021)\citenamefont
  {Abramowicz} \emph {et~al.}}]{Abramowicz:2021zja}%
  \BibitemOpen
  \bibfield  {author} {\bibinfo {author} {\bibfnamefont {H.}~\bibnamefont
  {Abramowicz}} \emph {et~al.},\ }\href {\doibase
  10.1140/epjs/s11734-021-00249-z} {\bibfield  {journal} {\bibinfo  {journal}
  {Eur. Phys. J. ST}\ }\textbf {\bibinfo {volume} {230}},\ \bibinfo {pages}
  {2445} (\bibinfo {year} {2021})},\ \Eprint {http://arxiv.org/abs/2102.02032}
  {arXiv:2102.02032 [hep-ex]} \BibitemShut {NoStop}%
\bibitem [{\citenamefont {Bula}(1997)}]{bula97}%
  \BibitemOpen
  \bibfield  {author} {\bibinfo {author} {\bibfnamefont {C.}~\bibnamefont
  {Bula}},\ }\href@noop {} {\enquote {\bibinfo {title} {{A numeric integration
  program to simulate nonlinear QED processes in Electron-laser or Photon-laser
  collisions}},}\ }\bibinfo {howpublished}
  {https://www.slac.stanford.edu/exp/e144/ps/nidoc.ps} (\bibinfo {year}
  {1997})\BibitemShut {NoStop}%
\bibitem [{\citenamefont {Bamber}\ \emph {et~al.}(1999)\citenamefont {Bamber},
  \citenamefont {Boege}, \citenamefont {Koffas}, \citenamefont {Kotseroglou},
  \citenamefont {Melissinos}, \citenamefont {Meyerhofer}, \citenamefont {Reis},
  \citenamefont {Ragg}, \citenamefont {Bula}, \citenamefont {McDonald},
  \citenamefont {Prebys}, \citenamefont {Burke}, \citenamefont {Field},
  \citenamefont {Horton-Smith}, \citenamefont {Spencer}, \citenamefont {Walz},
  \citenamefont {Berridge}, \citenamefont {Bugg}, \citenamefont {Shmakov},\
  and\ \citenamefont {Weidemann}}]{bamber.prd.1999}%
  \BibitemOpen
  \bibfield  {author} {\bibinfo {author} {\bibfnamefont {C.}~\bibnamefont
  {Bamber}}, \bibinfo {author} {\bibfnamefont {S.~J.}\ \bibnamefont {Boege}},
  \bibinfo {author} {\bibfnamefont {T.}~\bibnamefont {Koffas}}, \bibinfo
  {author} {\bibfnamefont {T.}~\bibnamefont {Kotseroglou}}, \bibinfo {author}
  {\bibfnamefont {A.~C.}\ \bibnamefont {Melissinos}}, \bibinfo {author}
  {\bibfnamefont {D.~D.}\ \bibnamefont {Meyerhofer}}, \bibinfo {author}
  {\bibfnamefont {D.~A.}\ \bibnamefont {Reis}}, \bibinfo {author}
  {\bibfnamefont {W.}~\bibnamefont {Ragg}}, \bibinfo {author} {\bibfnamefont
  {C.}~\bibnamefont {Bula}}, \bibinfo {author} {\bibfnamefont {K.~T.}\
  \bibnamefont {McDonald}}, \bibinfo {author} {\bibfnamefont {E.~J.}\
  \bibnamefont {Prebys}}, \bibinfo {author} {\bibfnamefont {D.~L.}\
  \bibnamefont {Burke}}, \bibinfo {author} {\bibfnamefont {R.~C.}\ \bibnamefont
  {Field}}, \bibinfo {author} {\bibfnamefont {G.}~\bibnamefont {Horton-Smith}},
  \bibinfo {author} {\bibfnamefont {J.~E.}\ \bibnamefont {Spencer}}, \bibinfo
  {author} {\bibfnamefont {D.}~\bibnamefont {Walz}}, \bibinfo {author}
  {\bibfnamefont {S.~C.}\ \bibnamefont {Berridge}}, \bibinfo {author}
  {\bibfnamefont {W.~M.}\ \bibnamefont {Bugg}}, \bibinfo {author}
  {\bibfnamefont {K.}~\bibnamefont {Shmakov}}, \ and\ \bibinfo {author}
  {\bibfnamefont {A.~W.}\ \bibnamefont {Weidemann}},\ }\href {\doibase
  10.1103/PhysRevD.60.092004} {\bibfield  {journal} {\bibinfo  {journal} {Phys.
  Rev. D}\ }\textbf {\bibinfo {volume} {60}},\ \bibinfo {pages} {092004}
  (\bibinfo {year} {1999})}\BibitemShut {NoStop}%
\bibitem [{\citenamefont {Heinzl}\ \emph {et~al.}(2020)\citenamefont {Heinzl},
  \citenamefont {King},\ and\ \citenamefont {Macleod}}]{Heinzl:2020ynb}%
  \BibitemOpen
  \bibfield  {author} {\bibinfo {author} {\bibfnamefont {T.}~\bibnamefont
  {Heinzl}}, \bibinfo {author} {\bibfnamefont {B.}~\bibnamefont {King}}, \ and\
  \bibinfo {author} {\bibfnamefont {A.~J.}\ \bibnamefont {Macleod}},\ }\href
  {\doibase 10.1103/PhysRevA.102.063110} {\bibfield  {journal} {\bibinfo
  {journal} {Phys. Rev. A}\ }\textbf {\bibinfo {volume} {102}},\ \bibinfo
  {pages} {063110} (\bibinfo {year} {2020})},\ \Eprint
  {http://arxiv.org/abs/2004.13035} {arXiv:2004.13035 [hep-ph]} \BibitemShut
  {NoStop}%
\bibitem [{\citenamefont {King}(2021)}]{King:2020hsk}%
  \BibitemOpen
  \bibfield  {author} {\bibinfo {author} {\bibfnamefont {B.}~\bibnamefont
  {King}},\ }\href {\doibase 10.1103/PhysRevD.103.036018} {\bibfield  {journal}
  {\bibinfo  {journal} {Phys. Rev. D}\ }\textbf {\bibinfo {volume} {103}},\
  \bibinfo {pages} {036018} (\bibinfo {year} {2021})},\ \Eprint
  {http://arxiv.org/abs/2012.05920} {arXiv:2012.05920 [hep-ph]} \BibitemShut
  {NoStop}%
\bibitem [{\citenamefont {Tang}\ and\ \citenamefont
  {King}(2021)}]{Tang:2021qht}%
  \BibitemOpen
  \bibfield  {author} {\bibinfo {author} {\bibfnamefont {S.}~\bibnamefont
  {Tang}}\ and\ \bibinfo {author} {\bibfnamefont {B.}~\bibnamefont {King}},\
  }\href {\doibase 10.1103/PhysRevD.104.096019} {\bibfield  {journal} {\bibinfo
   {journal} {Phys. Rev. D}\ }\textbf {\bibinfo {volume} {104}},\ \bibinfo
  {pages} {096019} (\bibinfo {year} {2021})},\ \Eprint
  {http://arxiv.org/abs/2109.00555} {arXiv:2109.00555 [physics.optics]}
  \BibitemShut {NoStop}%
\bibitem [{cai(1995)}]{cain}%
  \BibitemOpen
  \href {\doibase 10.1016/0168-9002(94)01186-9} {\bibfield  {journal} {\bibinfo
   {journal} {Nucl. Instrum. Methods Phys. Res. A}\ }\textbf {\bibinfo {volume}
  {355}},\ \bibinfo {pages} {107} (\bibinfo {year} {1995})}\BibitemShut
  {NoStop}%
\bibitem [{\citenamefont {Hartin}(2018)}]{hartin.ijmpa.2018}%
  \BibitemOpen
  \bibfield  {author} {\bibinfo {author} {\bibfnamefont {A.}~\bibnamefont
  {Hartin}},\ }\href {\doibase 10.1142/S0217751X18300119} {\bibfield  {journal}
  {\bibinfo  {journal} {Int. J. Mod. Phys. A}\ }\textbf {\bibinfo {volume}
  {33}},\ \bibinfo {pages} {1830011} (\bibinfo {year} {2018})}\BibitemShut
  {NoStop}%
\bibitem [{\citenamefont {Blackburn}(2022)}]{ptarmigan22}%
  \BibitemOpen
  \bibfield  {author} {\bibinfo {author} {\bibfnamefont {T.~G.}\ \bibnamefont
  {Blackburn}},\ }\href {https://github.com/tgblackburn/ptarmigan} {\enquote
  {\bibinfo {title} {\textsc{ptarmigan}},}\ } (\bibinfo {year}
  {2022})\BibitemShut {NoStop}%
\bibitem [{\citenamefont {Blackburn}\ and\ \citenamefont
  {King}(2022)}]{Blackburn:2021cuq}%
  \BibitemOpen
  \bibfield  {author} {\bibinfo {author} {\bibfnamefont {T.~G.}\ \bibnamefont
  {Blackburn}}\ and\ \bibinfo {author} {\bibfnamefont {B.}~\bibnamefont
  {King}},\ }\href {\doibase 10.1140/epjc/s10052-021-09955-3} {\bibfield
  {journal} {\bibinfo  {journal} {Eur. Phys. J. C}\ }\textbf {\bibinfo {volume}
  {82}},\ \bibinfo {pages} {44} (\bibinfo {year} {2022})},\ \Eprint
  {http://arxiv.org/abs/2108.10883} {arXiv:2108.10883 [hep-ph]} \BibitemShut
  {NoStop}%
\bibitem [{\citenamefont {Blackburn}\ \emph {et~al.}(2021)\citenamefont
  {Blackburn}, \citenamefont {MacLeod},\ and\ \citenamefont
  {King}}]{Blackburn:2021rqm}%
  \BibitemOpen
  \bibfield  {author} {\bibinfo {author} {\bibfnamefont {T.~G.}\ \bibnamefont
  {Blackburn}}, \bibinfo {author} {\bibfnamefont {A.~J.}\ \bibnamefont
  {MacLeod}}, \ and\ \bibinfo {author} {\bibfnamefont {B.}~\bibnamefont
  {King}},\ }\href {\doibase 10.1088/1367-2630/ac1bf6} {\bibfield  {journal}
  {\bibinfo  {journal} {New J. Phys.}\ }\textbf {\bibinfo {volume} {23}},\
  \bibinfo {pages} {085008} (\bibinfo {year} {2021})},\ \Eprint
  {http://arxiv.org/abs/2103.06673} {arXiv:2103.06673 [hep-ph]} \BibitemShut
  {NoStop}%
\bibitem [{\citenamefont {Tang}(2022)}]{Tang:2022tmn}%
  \BibitemOpen
  \bibfield  {author} {\bibinfo {author} {\bibfnamefont {S.}~\bibnamefont
  {Tang}},\ }\href@noop {} {\  (\bibinfo {year} {2022})},\ \Eprint
  {http://arxiv.org/abs/2203.05721} {arXiv:2203.05721 [hep-ph]} \BibitemShut
  {NoStop}%
\bibitem [{\citenamefont {Narozhnyi}\ and\ \citenamefont
  {Fofanov}(1996)}]{Narozhnyi:1996qf}%
  \BibitemOpen
  \bibfield  {author} {\bibinfo {author} {\bibfnamefont {N.~B.}\ \bibnamefont
  {Narozhnyi}}\ and\ \bibinfo {author} {\bibfnamefont {M.~S.}\ \bibnamefont
  {Fofanov}},\ }\href@noop {} {\bibfield  {journal} {\bibinfo  {journal} {J.
  Exp. Theor. Phys.}\ }\textbf {\bibinfo {volume} {83}},\ \bibinfo {pages} {14}
  (\bibinfo {year} {1996})},\ \bibinfo {note} {[Zh.\ Eksp.\ Teor.\ Fiz.\
  \textbf{110}, 26 (1996)]}\BibitemShut {NoStop}%
\bibitem [{\citenamefont {McDonald}(1997)}]{McDonald:1997}%
  \BibitemOpen
  \bibfield  {author} {\bibinfo {author} {\bibfnamefont {K.}~\bibnamefont
  {McDonald}},\ }\href {www.hep.princeton.edu/~mcdonald/examples/gaussian2.pdf}
  {\enquote {\bibinfo {title} {A relativistic electron can't extract net energy
  from a `long' laser pulse},}\ } (\bibinfo {year} {1997})\BibitemShut
  {NoStop}%
\bibitem [{\citenamefont {Seipt}\ and\ \citenamefont
  {K\"ampfer}(2011)}]{seipt.pra.2011}%
  \BibitemOpen
  \bibfield  {author} {\bibinfo {author} {\bibfnamefont {D.}~\bibnamefont
  {Seipt}}\ and\ \bibinfo {author} {\bibfnamefont {B.}~\bibnamefont
  {K\"ampfer}},\ }\href {\doibase 10.1103/PhysRevA.83.022101} {\bibfield
  {journal} {\bibinfo  {journal} {Phys. Rev. A}\ }\textbf {\bibinfo {volume}
  {83}},\ \bibinfo {pages} {022101} (\bibinfo {year} {2011})}\BibitemShut
  {NoStop}%
\bibitem [{\citenamefont {Seipt}\ \emph {et~al.}(2016)\citenamefont {Seipt},
  \citenamefont {Kharin}, \citenamefont {Rykovanov}, \citenamefont
  {Surzhykov},\ and\ \citenamefont {Fritzsche}}]{seipt.jpp.2016}%
  \BibitemOpen
  \bibfield  {author} {\bibinfo {author} {\bibfnamefont {D.}~\bibnamefont
  {Seipt}}, \bibinfo {author} {\bibfnamefont {V.}~\bibnamefont {Kharin}},
  \bibinfo {author} {\bibfnamefont {S.}~\bibnamefont {Rykovanov}}, \bibinfo
  {author} {\bibfnamefont {A.}~\bibnamefont {Surzhykov}}, \ and\ \bibinfo
  {author} {\bibfnamefont {S.}~\bibnamefont {Fritzsche}},\ }\href {\doibase
  0.1017/S002237781600026X} {\bibfield  {journal} {\bibinfo  {journal} {J.
  Plasma Phys.}\ }\textbf {\bibinfo {volume} {82}},\ \bibinfo {pages}
  {655820203} (\bibinfo {year} {2016})}\BibitemShut {NoStop}%
\bibitem [{\citenamefont {Baier}\ and\ \citenamefont
  {Katkov}(1968)}]{baier1968quasiclassical}%
  \BibitemOpen
  \bibfield  {author} {\bibinfo {author} {\bibfnamefont {V.}~\bibnamefont
  {Baier}}\ and\ \bibinfo {author} {\bibfnamefont {V.}~\bibnamefont {Katkov}},\
  }\href@noop {} {\bibfield  {journal} {\bibinfo  {journal} {Sov. Phys. JETP}\
  }\textbf {\bibinfo {volume} {26}},\ \bibinfo {pages} {854} (\bibinfo {year}
  {1968})}\BibitemShut {NoStop}%
\bibitem [{\citenamefont {Matveev}(1957)}]{Matveev1957}%
  \BibitemOpen
  \bibfield  {author} {\bibinfo {author} {\bibfnamefont {A.~N.}\ \bibnamefont
  {Matveev}},\ }\href@noop {} {\bibfield  {journal} {\bibinfo  {journal} {Sov.
  Phys. JETP}\ }\textbf {\bibinfo {volume} {4}},\ \bibinfo {pages} {409}
  (\bibinfo {year} {1957})}\BibitemShut {NoStop}%
\bibitem [{\citenamefont {Kimball}\ \emph {et~al.}(1986)\citenamefont
  {Kimball}, \citenamefont {Cue},\ and\ \citenamefont
  {Belkacem}}]{kimball_1986}%
  \BibitemOpen
  \bibfield  {author} {\bibinfo {author} {\bibfnamefont {J.}~\bibnamefont
  {Kimball}}, \bibinfo {author} {\bibfnamefont {N.}~\bibnamefont {Cue}}, \ and\
  \bibinfo {author} {\bibfnamefont {A.}~\bibnamefont {Belkacem}},\ }\href
  {\doibase https://doi.org/10.1016/0168-583X(86)90461-1} {\bibfield  {journal}
  {\bibinfo  {journal} {Nuclear Instruments and Methods in Physics Research
  Section B: Beam Interactions with Materials and Atoms}\ }\textbf {\bibinfo
  {volume} {13}},\ \bibinfo {pages} {1 } (\bibinfo {year} {1986})}\BibitemShut
  {NoStop}%
\bibitem [{\citenamefont {Di~Piazza}(2016)}]{DiPiazza:2016maj}%
  \BibitemOpen
  \bibfield  {author} {\bibinfo {author} {\bibfnamefont {A.}~\bibnamefont
  {Di~Piazza}},\ }\href {\doibase 10.1103/PhysRevLett.117.213201} {\bibfield
  {journal} {\bibinfo  {journal} {Phys. Rev. Lett.}\ }\textbf {\bibinfo
  {volume} {117}},\ \bibinfo {pages} {213201} (\bibinfo {year} {2016})},\
  \Eprint {http://arxiv.org/abs/1608.08120} {arXiv:1608.08120 [hep-ph]}
  \BibitemShut {NoStop}%
\bibitem [{\citenamefont {Di~Piazza}(2017)}]{DiPiazza:2016tdf}%
  \BibitemOpen
  \bibfield  {author} {\bibinfo {author} {\bibfnamefont {A.}~\bibnamefont
  {Di~Piazza}},\ }\href {\doibase 10.1103/PhysRevA.95.032121} {\bibfield
  {journal} {\bibinfo  {journal} {Phys. Rev. A}\ }\textbf {\bibinfo {volume}
  {95}},\ \bibinfo {pages} {032121} (\bibinfo {year} {2017})},\ \Eprint
  {http://arxiv.org/abs/1612.04132} {arXiv:1612.04132 [hep-ph]} \BibitemShut
  {NoStop}%
\bibitem [{\citenamefont {Di~Piazza}(2021)}]{DiPiazza:2020wxp}%
  \BibitemOpen
  \bibfield  {author} {\bibinfo {author} {\bibfnamefont {A.}~\bibnamefont
  {Di~Piazza}},\ }\href {\doibase 10.1103/PhysRevA.103.012215} {\bibfield
  {journal} {\bibinfo  {journal} {Phys. Rev. A}\ }\textbf {\bibinfo {volume}
  {103}},\ \bibinfo {pages} {012215} (\bibinfo {year} {2021})},\ \Eprint
  {http://arxiv.org/abs/2009.00526} {arXiv:2009.00526 [hep-ph]} \BibitemShut
  {NoStop}%
\bibitem [{\citenamefont {Di~Piazza}(2014)}]{DiPiazza:2013vra}%
  \BibitemOpen
  \bibfield  {author} {\bibinfo {author} {\bibfnamefont {A.}~\bibnamefont
  {Di~Piazza}},\ }\href {\doibase 10.1103/PhysRevLett.113.040402} {\bibfield
  {journal} {\bibinfo  {journal} {Phys. Rev. Lett.}\ }\textbf {\bibinfo
  {volume} {113}},\ \bibinfo {pages} {040402} (\bibinfo {year} {2014})},\
  \Eprint {http://arxiv.org/abs/1310.7856} {arXiv:1310.7856 [hep-ph]}
  \BibitemShut {NoStop}%
\bibitem [{\citenamefont {Lv}\ \emph {et~al.}(2021)\citenamefont {Lv},
  \citenamefont {Raicher}, \citenamefont {Keitel},\ and\ \citenamefont
  {Hatsagortsyan}}]{Lv:2021ayt}%
  \BibitemOpen
  \bibfield  {author} {\bibinfo {author} {\bibfnamefont {Q.~Z.}\ \bibnamefont
  {Lv}}, \bibinfo {author} {\bibfnamefont {E.}~\bibnamefont {Raicher}},
  \bibinfo {author} {\bibfnamefont {C.~H.}\ \bibnamefont {Keitel}}, \ and\
  \bibinfo {author} {\bibfnamefont {K.~Z.}\ \bibnamefont {Hatsagortsyan}},\
  }\href {\doibase 10.1088/1367-2630/abfa60} {\bibfield  {journal} {\bibinfo
  {journal} {New J. Phys.}\ }\textbf {\bibinfo {volume} {23}},\ \bibinfo
  {pages} {065005} (\bibinfo {year} {2021})},\ \Eprint
  {http://arxiv.org/abs/2106.01303} {arXiv:2106.01303 [physics.plasm-ph]}
  \BibitemShut {NoStop}%
\bibitem [{\citenamefont {Nielsen}\ \emph {et~al.}(2020)\citenamefont
  {Nielsen}, \citenamefont {Justesen}, \citenamefont {S\o{}rensen},
  \citenamefont {Uggerh\o{}j},\ and\ \citenamefont
  {Holtzapple}}]{Nielsen2020LL}%
  \BibitemOpen
  \bibfield  {author} {\bibinfo {author} {\bibfnamefont {C.~F.}\ \bibnamefont
  {Nielsen}}, \bibinfo {author} {\bibfnamefont {J.~B.}\ \bibnamefont
  {Justesen}}, \bibinfo {author} {\bibfnamefont {A.~H.}\ \bibnamefont
  {S\o{}rensen}}, \bibinfo {author} {\bibfnamefont {U.~I.}\ \bibnamefont
  {Uggerh\o{}j}}, \ and\ \bibinfo {author} {\bibfnamefont {R.}~\bibnamefont
  {Holtzapple}} (\bibinfo {collaboration} {CERN NA63 Collaboration}),\ }\href
  {\doibase 10.1103/PhysRevD.102.052004} {\bibfield  {journal} {\bibinfo
  {journal} {Phys. Rev. D}\ }\textbf {\bibinfo {volume} {102}},\ \bibinfo
  {pages} {052004} (\bibinfo {year} {2020})}\BibitemShut {NoStop}%
\bibitem [{\citenamefont {Nielsen}(2020)}]{NielsenGPU2020}%
  \BibitemOpen
  \bibfield  {author} {\bibinfo {author} {\bibfnamefont {C.}~\bibnamefont
  {Nielsen}},\ }\href {\doibase 10.1016/j.cpc.2019.107128} {\bibfield
  {journal} {\bibinfo  {journal} {Computer Physics Communications}\ }\textbf
  {\bibinfo {volume} {252}} (\bibinfo {year} {2020}),\
  10.1016/j.cpc.2019.107128}\BibitemShut {NoStop}%
\bibitem [{\citenamefont {Berestetskii}\ \emph {et~al.}(1982)\citenamefont
  {Berestetskii}, \citenamefont {Lifshitz},\ and\ \citenamefont
  {Pitaevskii}}]{BERESTETSKII1982}%
  \BibitemOpen
  \bibfield  {author} {\bibinfo {author} {\bibfnamefont {V.}~\bibnamefont
  {Berestetskii}}, \bibinfo {author} {\bibfnamefont {E.}~\bibnamefont
  {Lifshitz}}, \ and\ \bibinfo {author} {\bibfnamefont {L.}~\bibnamefont
  {Pitaevskii}},\ }in\ \href {\doibase
  https://doi.org/10.1016/B978-0-08-050346-2.50012-X} {\emph {\bibinfo
  {booktitle} {Quantum Electrodynamics}}}\ (\bibinfo  {publisher}
  {Butterworth-Heinemann},\ \bibinfo {address} {Oxford},\ \bibinfo {year}
  {1982})\ \bibinfo {edition} {2nd}\ ed.\BibitemShut {Stop}%
\bibitem [{\citenamefont {Volkov}(1935)}]{volkov1935class}%
  \BibitemOpen
  \bibfield  {author} {\bibinfo {author} {\bibfnamefont {D.}~\bibnamefont
  {Volkov}},\ }\href@noop {} {\bibfield  {journal} {\bibinfo  {journal} {Z.
  Phys}\ }\textbf {\bibinfo {volume} {94}},\ \bibinfo {pages} {250} (\bibinfo
  {year} {1935})}\BibitemShut {NoStop}%
\bibitem [{\citenamefont {Jackson}(1998)}]{jackson_classical_1998}%
  \BibitemOpen
  \bibfield  {author} {\bibinfo {author} {\bibfnamefont {J.~D.}\ \bibnamefont
  {Jackson}},\ }\href@noop {} {\emph {\bibinfo {title} {Classical
  Electrodynamics}}},\ \bibinfo {edition} {3rd}\ ed.\ (\bibinfo  {publisher}
  {John Wiley \& Sons},\ \bibinfo {year} {1998})\BibitemShut {NoStop}%
\bibitem [{\citenamefont {Blackburn}\ \emph {et~al.}(2014)\citenamefont
  {Blackburn}, \citenamefont {Ridgers}, \citenamefont {Kirk},\ and\
  \citenamefont {Bell}}]{Blackburn:2014cig}%
  \BibitemOpen
  \bibfield  {author} {\bibinfo {author} {\bibfnamefont {T.~G.}\ \bibnamefont
  {Blackburn}}, \bibinfo {author} {\bibfnamefont {C.~P.}\ \bibnamefont
  {Ridgers}}, \bibinfo {author} {\bibfnamefont {J.~G.}\ \bibnamefont {Kirk}}, \
  and\ \bibinfo {author} {\bibfnamefont {A.~R.}\ \bibnamefont {Bell}},\ }\href
  {\doibase 10.1103/PhysRevLett.112.015001} {\bibfield  {journal} {\bibinfo
  {journal} {Phys. Rev. Lett.}\ }\textbf {\bibinfo {volume} {112}},\ \bibinfo
  {pages} {015001} (\bibinfo {year} {2014})},\ \Eprint
  {http://arxiv.org/abs/1503.01009} {arXiv:1503.01009 [physics.plasm-ph]}
  \BibitemShut {NoStop}%
\bibitem [{\citenamefont {Harvey}\ \emph {et~al.}(2017)\citenamefont {Harvey},
  \citenamefont {Gonoskov}, \citenamefont {Ilderton},\ and\ \citenamefont
  {Marklund}}]{Harvey:2016uiy}%
  \BibitemOpen
  \bibfield  {author} {\bibinfo {author} {\bibfnamefont {C.}~\bibnamefont
  {Harvey}}, \bibinfo {author} {\bibfnamefont {A.}~\bibnamefont {Gonoskov}},
  \bibinfo {author} {\bibfnamefont {A.}~\bibnamefont {Ilderton}}, \ and\
  \bibinfo {author} {\bibfnamefont {M.}~\bibnamefont {Marklund}},\ }\href
  {\doibase 10.1103/PhysRevLett.118.105004} {\bibfield  {journal} {\bibinfo
  {journal} {Phys. Rev. Lett.}\ }\textbf {\bibinfo {volume} {118}},\ \bibinfo
  {pages} {105004} (\bibinfo {year} {2017})},\ \Eprint
  {http://arxiv.org/abs/1606.08250} {arXiv:1606.08250 [hep-ph]} \BibitemShut
  {NoStop}%
\bibitem [{\citenamefont {Nikishov}\ and\ \citenamefont
  {Ritus}(1964)}]{Nikishov1964}%
  \BibitemOpen
  \bibfield  {author} {\bibinfo {author} {\bibfnamefont {A.~I.}\ \bibnamefont
  {Nikishov}}\ and\ \bibinfo {author} {\bibfnamefont {V.~I.}\ \bibnamefont
  {Ritus}},\ }\href@noop {} {\bibfield  {journal} {\bibinfo  {journal} {Sov.
  Phys. JETP}\ }\textbf {\bibinfo {volume} {19}},\ \bibinfo {pages} {529}
  (\bibinfo {year} {1964})}\BibitemShut {NoStop}%
\bibitem [{\citenamefont {Brown}\ and\ \citenamefont
  {Kibble}(1964)}]{Brown:1964zzb}%
  \BibitemOpen
  \bibfield  {author} {\bibinfo {author} {\bibfnamefont {L.~S.}\ \bibnamefont
  {Brown}}\ and\ \bibinfo {author} {\bibfnamefont {T.~W.~B.}\ \bibnamefont
  {Kibble}},\ }\href {\doibase 10.1103/PhysRev.133.A705} {\bibfield  {journal}
  {\bibinfo  {journal} {Phys. Rev.}\ }\textbf {\bibinfo {volume} {133}},\
  \bibinfo {pages} {A705} (\bibinfo {year} {1964})}\BibitemShut {NoStop}%
\bibitem [{\citenamefont {L\"otstedt}\ and\ \citenamefont
  {Jentschura}(2009)}]{PhysRevE.79.026707}%
  \BibitemOpen
  \bibfield  {author} {\bibinfo {author} {\bibfnamefont {E.}~\bibnamefont
  {L\"otstedt}}\ and\ \bibinfo {author} {\bibfnamefont {U.~D.}\ \bibnamefont
  {Jentschura}},\ }\href {\doibase 10.1103/PhysRevE.79.026707} {\bibfield
  {journal} {\bibinfo  {journal} {Phys. Rev. E}\ }\textbf {\bibinfo {volume}
  {79}},\ \bibinfo {pages} {026707} (\bibinfo {year} {2009})}\BibitemShut
  {NoStop}%
\bibitem [{\citenamefont {Bula}\ \emph
  {et~al.}(1996{\natexlab{b}})\citenamefont {Bula}, \citenamefont {McDonald},
  \citenamefont {Prebys}, \citenamefont {Bamber}, \citenamefont {Boege},
  \citenamefont {Kotseroglou}, \citenamefont {Melissinos}, \citenamefont
  {Meyerhofer}, \citenamefont {Ragg}, \citenamefont {Burke}, \citenamefont
  {Field}, \citenamefont {Horton-Smith}, \citenamefont {Odian}, \citenamefont
  {Spencer}, \citenamefont {Walz}, \citenamefont {Berridge}, \citenamefont
  {Bugg}, \citenamefont {Shmakov},\ and\ \citenamefont {Weidemann}}]{E144}%
  \BibitemOpen
  \bibfield  {author} {\bibinfo {author} {\bibfnamefont {C.}~\bibnamefont
  {Bula}}, \bibinfo {author} {\bibfnamefont {K.~T.}\ \bibnamefont {McDonald}},
  \bibinfo {author} {\bibfnamefont {E.~J.}\ \bibnamefont {Prebys}}, \bibinfo
  {author} {\bibfnamefont {C.}~\bibnamefont {Bamber}}, \bibinfo {author}
  {\bibfnamefont {S.}~\bibnamefont {Boege}}, \bibinfo {author} {\bibfnamefont
  {T.}~\bibnamefont {Kotseroglou}}, \bibinfo {author} {\bibfnamefont {A.~C.}\
  \bibnamefont {Melissinos}}, \bibinfo {author} {\bibfnamefont {D.~D.}\
  \bibnamefont {Meyerhofer}}, \bibinfo {author} {\bibfnamefont
  {W.}~\bibnamefont {Ragg}}, \bibinfo {author} {\bibfnamefont {D.~L.}\
  \bibnamefont {Burke}}, \bibinfo {author} {\bibfnamefont {R.~C.}\ \bibnamefont
  {Field}}, \bibinfo {author} {\bibfnamefont {G.}~\bibnamefont {Horton-Smith}},
  \bibinfo {author} {\bibfnamefont {A.~C.}\ \bibnamefont {Odian}}, \bibinfo
  {author} {\bibfnamefont {J.~E.}\ \bibnamefont {Spencer}}, \bibinfo {author}
  {\bibfnamefont {D.}~\bibnamefont {Walz}}, \bibinfo {author} {\bibfnamefont
  {S.~C.}\ \bibnamefont {Berridge}}, \bibinfo {author} {\bibfnamefont {W.~M.}\
  \bibnamefont {Bugg}}, \bibinfo {author} {\bibfnamefont {K.}~\bibnamefont
  {Shmakov}}, \ and\ \bibinfo {author} {\bibfnamefont {A.~W.}\ \bibnamefont
  {Weidemann}},\ }\href {\doibase 10.1103/PhysRevLett.76.3116} {\bibfield
  {journal} {\bibinfo  {journal} {Phys. Rev. Lett.}\ }\textbf {\bibinfo
  {volume} {76}},\ \bibinfo {pages} {3116} (\bibinfo {year}
  {1996}{\natexlab{b}})}\BibitemShut {NoStop}%
\bibitem [{\citenamefont {Khokonov}\ and\ \citenamefont
  {Nitta}(2002)}]{PhysRevLett.89.094801}%
  \BibitemOpen
  \bibfield  {author} {\bibinfo {author} {\bibfnamefont {M.~K.}\ \bibnamefont
  {Khokonov}}\ and\ \bibinfo {author} {\bibfnamefont {H.}~\bibnamefont
  {Nitta}},\ }\href {\doibase 10.1103/PhysRevLett.89.094801} {\bibfield
  {journal} {\bibinfo  {journal} {Phys. Rev. Lett.}\ }\textbf {\bibinfo
  {volume} {89}},\ \bibinfo {pages} {094801} (\bibinfo {year}
  {2002})}\BibitemShut {NoStop}%
\bibitem [{\citenamefont {Baier}\ \emph {et~al.}(1998)\citenamefont {Baier},
  \citenamefont {Katkov},\ and\ \citenamefont {Strakhovenko}}]{Baier}%
  \BibitemOpen
  \bibfield  {author} {\bibinfo {author} {\bibfnamefont {V.~N.}\ \bibnamefont
  {Baier}}, \bibinfo {author} {\bibfnamefont {V.~M.}\ \bibnamefont {Katkov}}, \
  and\ \bibinfo {author} {\bibfnamefont {V.~M.}\ \bibnamefont {Strakhovenko}},\
  }\href@noop {} {\emph {\bibinfo {title} {{Electromagnetic processes at high
  energies in oriented single crystals}}}}\ (\bibinfo {year}
  {1998})\BibitemShut {NoStop}%
\bibitem [{\citenamefont {Belkacem}\ \emph {et~al.}(1985)\citenamefont
  {Belkacem}, \citenamefont {Cue},\ and\ \citenamefont
  {Kimball}}]{belkacem_1985}%
  \BibitemOpen
  \bibfield  {author} {\bibinfo {author} {\bibfnamefont {A.}~\bibnamefont
  {Belkacem}}, \bibinfo {author} {\bibfnamefont {N.}~\bibnamefont {Cue}}, \
  and\ \bibinfo {author} {\bibfnamefont {J.}~\bibnamefont {Kimball}},\ }\href
  {\doibase https://doi.org/10.1016/0375-9601(85)90811-4} {\bibfield  {journal}
  {\bibinfo  {journal} {Physics Letters A}\ }\textbf {\bibinfo {volume}
  {111}},\ \bibinfo {pages} {86 } (\bibinfo {year} {1985})}\BibitemShut
  {NoStop}%
\bibitem [{\citenamefont {Wistisen}(2014{\natexlab{a}})}]{TobiasUdvikling}%
  \BibitemOpen
  \bibfield  {author} {\bibinfo {author} {\bibfnamefont {T.~N.}\ \bibnamefont
  {Wistisen}},\ }\href {https://link.aps.org/doi/10.1103/PhysRevD.90.125008}
  {\bibfield  {journal} {\bibinfo  {journal} {Phys. Rev. D}\ }\textbf {\bibinfo
  {volume} {90}},\ \bibinfo {pages} {125008} (\bibinfo {year}
  {2014}{\natexlab{a}})}\BibitemShut {NoStop}%
\bibitem [{\citenamefont {Lindhard}(1991)}]{Lind91}%
  \BibitemOpen
  \bibfield  {author} {\bibinfo {author} {\bibfnamefont {J.}~\bibnamefont
  {Lindhard}},\ }\href {\doibase 10.1103/PhysRevA.43.6032} {\bibfield
  {journal} {\bibinfo  {journal} {Phys. Rev. A}\ }\textbf {\bibinfo {volume}
  {43}},\ \bibinfo {pages} {6032} (\bibinfo {year} {1991})}\BibitemShut
  {NoStop}%
\bibitem [{\citenamefont {Wistisen}\ \emph {et~al.}(2019)\citenamefont
  {Wistisen}, \citenamefont {Piazza}, \citenamefont {Nielsen}, \citenamefont
  {S{\o}rensen},\ and\ \citenamefont {Uggerh{\o}j}}]{Tobias2019}%
  \BibitemOpen
  \bibfield  {author} {\bibinfo {author} {\bibfnamefont {T.}~\bibnamefont
  {Wistisen}}, \bibinfo {author} {\bibfnamefont {A.}~\bibnamefont {Piazza}},
  \bibinfo {author} {\bibfnamefont {C.}~\bibnamefont {Nielsen}}, \bibinfo
  {author} {\bibfnamefont {A.}~\bibnamefont {S{\o}rensen}}, \ and\ \bibinfo
  {author} {\bibfnamefont {U.}~\bibnamefont {Uggerh{\o}j}},\ }\href {\doibase
  10.1103/PhysRevResearch.1.033014} {\bibfield  {journal} {\bibinfo  {journal}
  {Physical Review Research}\ }\textbf {\bibinfo {volume} {1}} (\bibinfo {year}
  {2019}),\ 10.1103/PhysRevResearch.1.033014}\BibitemShut {NoStop}%
\bibitem [{\citenamefont {Tanabashi~{\em{et al.\/}}}(2018)}]{PDG_2018}%
  \BibitemOpen
  \bibfield  {author} {\bibinfo {author} {\bibfnamefont {M.}~\bibnamefont
  {Tanabashi~{\em{et al.\/}}}} (\bibinfo {collaboration} {Particle Data
  Group}),\ }\href {\doibase 10.1103/PhysRevD.98.030001} {\bibfield  {journal}
  {\bibinfo  {journal} {Phys. Rev. D}\ }\textbf {\bibinfo {volume} {98}},\
  \bibinfo {pages} {030001} (\bibinfo {year} {2018})}\BibitemShut {NoStop}%
\bibitem [{\citenamefont {Andersen}\ \emph {et~al.}(2012)\citenamefont
  {Andersen}, \citenamefont {Andersen}, \citenamefont {Esberg}, \citenamefont
  {Knudsen}, \citenamefont {Mikkelsen}, \citenamefont {Uggerh\o{}j},
  \citenamefont {Sona}, \citenamefont {Mangiarotti}, \citenamefont {Ketel},\
  and\ \citenamefont {Ballestrero}}]{UlrikFormation}%
  \BibitemOpen
  \bibfield  {author} {\bibinfo {author} {\bibfnamefont {K.~K.}\ \bibnamefont
  {Andersen}}, \bibinfo {author} {\bibfnamefont {S.~L.}\ \bibnamefont
  {Andersen}}, \bibinfo {author} {\bibfnamefont {J.}~\bibnamefont {Esberg}},
  \bibinfo {author} {\bibfnamefont {H.}~\bibnamefont {Knudsen}}, \bibinfo
  {author} {\bibfnamefont {R.}~\bibnamefont {Mikkelsen}}, \bibinfo {author}
  {\bibfnamefont {U.~I.}\ \bibnamefont {Uggerh\o{}j}}, \bibinfo {author}
  {\bibfnamefont {P.}~\bibnamefont {Sona}}, \bibinfo {author} {\bibfnamefont
  {A.}~\bibnamefont {Mangiarotti}}, \bibinfo {author} {\bibfnamefont {T.~J.}\
  \bibnamefont {Ketel}}, \ and\ \bibinfo {author} {\bibfnamefont
  {S.}~\bibnamefont {Ballestrero}} (\bibinfo {collaboration} {CERN NA63}),\
  }\href {\doibase 10.1103/PhysRevLett.108.071802} {\bibfield  {journal}
  {\bibinfo  {journal} {Phys. Rev. Lett.}\ }\textbf {\bibinfo {volume} {108}},\
  \bibinfo {pages} {071802} (\bibinfo {year} {2012})}\BibitemShut {NoStop}%
\bibitem [{\citenamefont {Wistisen}(2014{\natexlab{b}})}]{TobiasCompton}%
  \BibitemOpen
  \bibfield  {author} {\bibinfo {author} {\bibfnamefont {T.}~\bibnamefont
  {Wistisen}},\ }\href@noop {} {\bibfield  {journal} {\bibinfo  {journal}
  {Physical Review D}\ }\textbf {\bibinfo {volume} {90}} (\bibinfo {year}
  {2014}{\natexlab{b}})}\BibitemShut {NoStop}%
\bibitem [{\citenamefont {Titov}\ and\ \citenamefont
  {Kampfer}(2020)}]{Titov:2020taw}%
  \BibitemOpen
  \bibfield  {author} {\bibinfo {author} {\bibfnamefont {A.~I.}\ \bibnamefont
  {Titov}}\ and\ \bibinfo {author} {\bibfnamefont {B.}~\bibnamefont
  {Kampfer}},\ }\href {\doibase 10.1140/epjd/e2020-10327-9} {\bibfield
  {journal} {\bibinfo  {journal} {Eur. Phys. J. D}\ }\textbf {\bibinfo {volume}
  {74}},\ \bibinfo {pages} {218} (\bibinfo {year} {2020})},\ \Eprint
  {http://arxiv.org/abs/2006.04496} {arXiv:2006.04496 [hep-ph]} \BibitemShut
  {NoStop}%
\bibitem [{\citenamefont {Harvey}\ \emph
  {et~al.}(2015{\natexlab{b}})\citenamefont {Harvey}, \citenamefont
  {Ilderton},\ and\ \citenamefont {King}}]{harvey2015testing}%
  \BibitemOpen
  \bibfield  {author} {\bibinfo {author} {\bibfnamefont {C.}~\bibnamefont
  {Harvey}}, \bibinfo {author} {\bibfnamefont {A.}~\bibnamefont {Ilderton}}, \
  and\ \bibinfo {author} {\bibfnamefont {B.}~\bibnamefont {King}},\ }\href@noop
  {} {\bibfield  {journal} {\bibinfo  {journal} {Physical Review A}\ }\textbf
  {\bibinfo {volume} {91}},\ \bibinfo {pages} {013822} (\bibinfo {year}
  {2015}{\natexlab{b}})}\BibitemShut {NoStop}%
\bibitem [{\citenamefont {Siegman}(1986)}]{siegman_lasers_1986}%
  \BibitemOpen
  \bibfield  {author} {\bibinfo {author} {\bibfnamefont {A.~E.}\ \bibnamefont
  {Siegman}},\ }\href@noop {} {\emph {\bibinfo {title} {{L}asers}}}\ (\bibinfo
  {publisher} {University Science Books},\ \bibinfo {year} {1986})\BibitemShut
  {NoStop}%
\bibitem [{\citenamefont {Kogelnik}\ and\ \citenamefont
  {Li}(1966)}]{kogelnik_laser_1966}%
  \BibitemOpen
  \bibfield  {author} {\bibinfo {author} {\bibfnamefont {H.}~\bibnamefont
  {Kogelnik}}\ and\ \bibinfo {author} {\bibfnamefont {T.}~\bibnamefont {Li}},\
  }\href {\doibase 10.1364/AO.5.001550} {\bibfield  {journal} {\bibinfo
  {journal} {Appl. Opt.}\ }\textbf {\bibinfo {volume} {5}},\ \bibinfo {pages}
  {1550} (\bibinfo {year} {1966})}\BibitemShut {NoStop}%
\bibitem [{\citenamefont {Edwards}\ and\ \citenamefont
  {Ilderton}(2021)}]{Edwards:2020npu}%
  \BibitemOpen
  \bibfield  {author} {\bibinfo {author} {\bibfnamefont {J.~P.}\ \bibnamefont
  {Edwards}}\ and\ \bibinfo {author} {\bibfnamefont {A.}~\bibnamefont
  {Ilderton}},\ }\href {\doibase 10.1103/PhysRevD.103.016004} {\bibfield
  {journal} {\bibinfo  {journal} {Phys. Rev. D}\ }\textbf {\bibinfo {volume}
  {103}},\ \bibinfo {pages} {016004} (\bibinfo {year} {2021})},\ \Eprint
  {http://arxiv.org/abs/2010.02085} {arXiv:2010.02085 [hep-ph]} \BibitemShut
  {NoStop}%
\bibitem [{\citenamefont {Sengupta}(1952)}]{sengupta52}%
  \BibitemOpen
  \bibfield  {author} {\bibinfo {author} {\bibfnamefont {N.~D.}\ \bibnamefont
  {Sengupta}},\ }\href@noop {} {\bibfield  {journal} {\bibinfo  {journal}
  {Bull. Math. Soc. Calcutta}\ }\textbf {\bibinfo {volume} {44}} (\bibinfo
  {year} {1952})}\BibitemShut {NoStop}%
\bibitem [{\citenamefont {Harvey}\ \emph {et~al.}(2012)\citenamefont {Harvey},
  \citenamefont {Heinzl}, \citenamefont {Ilderton},\ and\ \citenamefont
  {Marklund}}]{Harvey:2012ie}%
  \BibitemOpen
  \bibfield  {author} {\bibinfo {author} {\bibfnamefont {C.}~\bibnamefont
  {Harvey}}, \bibinfo {author} {\bibfnamefont {T.}~\bibnamefont {Heinzl}},
  \bibinfo {author} {\bibfnamefont {A.}~\bibnamefont {Ilderton}}, \ and\
  \bibinfo {author} {\bibfnamefont {M.}~\bibnamefont {Marklund}},\ }\href
  {\doibase 10.1103/PhysRevLett.109.100402} {\bibfield  {journal} {\bibinfo
  {journal} {Phys. Rev. Lett.}\ }\textbf {\bibinfo {volume} {109}},\ \bibinfo
  {pages} {100402} (\bibinfo {year} {2012})},\ \Eprint
  {http://arxiv.org/abs/1203.6077} {arXiv:1203.6077 [hep-ph]} \BibitemShut
  {NoStop}%
\end{thebibliography}%

\end{document}